\documentclass[12pt,prd,aps,amssymb,amsmath,tightenlines,showpacs]{article}
\usepackage[utf8]{inputenc}
\usepackage{a4wide,amssymb,cite}
\pdfoutput=1

\usepackage{a4wide,amssymb,graphicx}
\usepackage{epsfig}
\usepackage{bm}
\usepackage{here}
\usepackage{cite}
\usepackage[usenames,dvipsnames]{color}
\usepackage{amssymb,cite,graphicx}
\usepackage{slashed}
\usepackage{amsmath,bm,bbm}
\usepackage{amsfonts}
\usepackage[titletoc,title]{appendix}
\usepackage[small]{caption}
\usepackage[margin=1in]{geometry}
\usepackage[multiple]{footmisc}
\usepackage{mathtools}
\usepackage{slashed}
\usepackage[nottoc]{tocbibind}
\usepackage{xcolor}
\usepackage{cancel}
\usepackage{comment}

 \usepackage{hyperref}
 \hypersetup{colorlinks,bookmarks,unicode,linktocpage=true,linkcolor=blue, anchorcolor=blue, citecolor=blue}

\newcommand{\be}{\begin{equation}}
	\newcommand{\ee}{\end{equation}}
\newcommand{\bea}{\begin{eqnarray}}
	\newcommand{\eea}{\end{eqnarray}}

\def\circa#1{\,\raise.3ex\hbox{$#1$\kern-.75em\lower1ex\hbox{$\sim$}}\,}

\begin{document}

\begin{titlepage}

\begin{centering}
\vspace{1cm}
{\Large \bf  Standard Model anomalies and vacuum stability \\ \vspace{0.2cm} for lepton portals with extra $U(1)$ symmetry } \\

\vspace{1.5cm}

{\bf Carlo Branchina$^{\ddagger,1,2}$, Hyun Min Lee$^{\dagger,2}$ and Kimiko Yamashita$^{\star,3}$  }
\vspace{.5cm}

{\it  $^1$Department of Physics, University of Calabria, I-87036 \\Arcavacata di Rende, Cosenza, Italy,\\
INFN-Cosenza, I-87036 Arcavacata di Rende, Cosenza, Italy.} 
\\[0.2cm]
{\it  $^2$Department of Physics, Chung-Ang University, Seoul 06974, Korea.} 
\\[0.2cm]
{\it $^3$Department of Physics, Ibaraki University, Mito 310-8512, Japan.}

\vspace{.5cm}
			

\end{centering}
\vspace{1.6cm}

\begin{abstract}
\noindent
Recently, the experimental values of the muon $(g-2)_\mu$ and of the $W$ boson mass $m_{_W}$ have both indicated significant deviations from the SM predictions, motivating the exploration of extensions with extra particles and symmetries. 
We revisit a lepton portal model with $U(1)'$ gauge symmetry where an extra Higgs doublet, a scalar singlet and one $SU(2)_L$ singlet vector-like fermion are introduced. In this model, $(g-2)_\mu$ can be explained by extra one-loop contributions from the vector-like lepton and the $Z'$ boson, whereas $m_{_W}$ can be increased by a tree-level mixing between the $Z$ and $Z'$.
Setting the $Z'$ and lepton couplings at low energies to account for the SM anomalies, we perform a Renormalization Group  analysis to investigate on the high-energy behaviour of the model, in particular on the issue of vacuum stability. We find that  in the alignment limit for the two Higgs doublets, the Landau pole and the scale where perturbativity is lost are of order $10-100\,{\rm TeV}$, not far from the scales experimentally reached so far, and sensibly lower than the stability scale. We show how the Landau pole can be increased up to $\sim10^9\,{\rm GeV}$ in a misaligned scenario where the experimental anomalies are still accommodated and a positive shift of the Higgs quartic coupling to improve stability can be achieved.

\end{abstract}

\vspace{2.5cm}

\begin{flushleft} 
$^\ddagger$Email: carlo.branchina@unical.it \\
$^\dagger$Email: hminlee@cau.ac.kr \\
$^\star$Email: kimiko.yamashita.nd93@vc.ibaraki.ac.jp
\end{flushleft}

\end{titlepage}
	
{\small
\tableofcontents
\newpage
}

\section{Introduction}

The measurements of the mass and couplings of the Higgs boson have turned out to be consistent with the Standard Model (SM) predictions.
Using the well-known SM relations, we can infer the Higgs quartic coupling $\lambda$ from the measured Higgs mass $m_H$ at low energy, and extrapolate it to higher energies through Renormalization Group (RG) equations. It has been shown that, within the SM, the running coupling $\lambda(\mu)$ turns negative at high energies due to the large top Yukawa coupling \cite{RGHiggs}, opening the possibility for a decay of the electroweak (EW) vacuum towards the true, deeper vacuum. This is at the origin of the vacuum stability problem in the SM, that might indicate the necessity for threshold corrections from new physics to appear below the stability scale \cite{threshold}, i.e. the scale where $\lambda(\mu)$ turns negative. Therefore, it is important to check how the picture changes in physics  Beyond the Standard Model (BSM) whenever the Higgs sector is extended or extra interactions for the SM Higgs boson are considered. New physics can impact the stability of the vacuum mainly in two ways. When new physics appears below the stability scale and it modifies the running of the couplings, this might even erase the possibility of a vacuum deeper than the EW one altogether. Even if the EW vacuum remains a false one, the tunneling time of the EW  vacuum can be subject to higher dimensional operators in the UV \cite{Branchina:2013jra}.

Recently, the measured values of the muon $(g-2)_\mu$ \cite{Muong-2:2006rrc,Muong-2:2021ojo} and the $W$ boson mass $m_{_W}$ \cite{CDF:2022hxs} have both indicated significant deviations from the SM predictions \cite{SM}, that we will refer to as the SM anomalies in the following, and have motivated the search for an explanation within extensions of the SM with extra particles and symmetries. However, it has been recently shown that the SM prediction  for the muon $g-2$ based on the lattice results is consistent with its experimental value \cite{lattice,lattice2}. Furthermore, the recent CMD-3  data  \cite{CMD-3} shows a sizable deviation from the other $e^+e^-$ data, which were used to derive the SM prediction for the muon $g-2$  in the dispersive approach taken in Ref.~\cite{SM}. Thus, it is important to understand the hadronic contributions to the muon $g-2$ well within the SM and there is a need of improvements on the experimental data. Nonetheless, it becomes of interest to investigate what the impact of new physics on the stability issue is in the phenomenologically motivated models as discussed above and provide more information from the complementary tests of new physics. 

In this article, we revisit a lepton portal model with an extra $U(1)'$ gauge symmetry from this perspective, and perform an RG analysis to investigate on its stability. In this model, suggested by some of the authors \cite{Lee:2022nqz}, an $SU(2)_L$ singlet vector-like lepton charged under $U(1)'$ is introduced. The vector-like lepton mixes with the muon through the vacuum expectation values (VEVs) of an extra Higgs doublet and a dark Higgs field, leading to a small seesaw mass for the muon. The muon $g-2$ can be explained thanks to extra one-loop corrections coming from the vector-lepton and the $Z'$ boson, while the $W$ mass $m_{_W}$ can be increased, at the same time, through a tree-level mixing between the $Z$ boson and the $Z'$ boson when the $Z'$ is heavier than $Z$. As the muon $g-2$ anomaly favors the $U(1)'$ breaking scale to be below about $200\,{\rm GeV}$, the extra gauge coupling required to increase $m_{_W}$  tends to be large \cite{Lee:2022nqz}, so that it can have a strong impact on the RG flow of the theory. There are previous analyses of RG equations and Higgs vacuum stability in the extended models with vector-like leptons and extended Higgs sectors with a motivation to explain the muon $g-2$ anomaly \cite{Hiller:2019mou,Hiller:2020fbu}. In our work, making a complete study of the one-loop RG equations in our model, we aim to go beyond the limited discussion on loop corrections done in the previous work in Ref.~\cite{Lee:2022nqz}. As a result, we present new results on the stability and the Landau pole of the model in the parameter space favored by both the muon $g-2$ and $W$ boson mass anomalies. However, in view of the recent lattice results for the muon $g-2$, it is notable that our RG analysis is not limited to the parameter space for the muon $g-2$ anomaly. 

Setting the couplings of the $Z'$ boson and of the vector-like lepton at low energy to explain the SM anomalies, we perform the RG analysis of the model. Following \cite{Lee:2022nqz}, we first consider the alignment limit of the Higgs potential by choosing some particular relations between the quartic couplings in the extended Higgs sector and identify the Landau pole and the scale where perturbativity is lost. We then extend the RG analysis to more general cases with ``minimal relations" between the quartic couplings, i.e.\,imposing only the relations needed to reproduce the experimental constraints and allowing for a misalignment between the CP-even neutral components of the Higgs doublets. The results of our analysis allow to determine a region of parameter space where the SM anomalies can be explained, the Landau pole of the theory is sufficiently higher than the scales experimentally probed so far, making it possible for the theory to have a UV cutoff at high enough scales, and the running of the quartic coupling can be lifted to positive values to (possibly) stabilize the vacuum.

The paper is organized as follows.
We present the model setup and the lepton portal interactions of the vector-like lepton in section 2.
We then consider the phenomenological constraints from the SM anomalies as well as $Z'$ searches at the Large Hadron Collider (LHC) in section 3.
Next we provide the analysis of vacuum stability and perturbativity with the running couplings in the model, focusing on the alignment limit of the extended Higgs potential and the low-energy inputs from the SM anomalies in section 4. We discuss the RG analysis for more general cases where the Higgs quartic couplings are deviated from the alignment limit and more general conditions for decoupling the dark Higgs boson are considered in section 5. 
There are two appendices where we present the calculation of the effective mass matrices for scalars, gauge bosons and fermions in this model and of the RG equations.
Finally, conclusions are drawn.

\section{The model}

Besides the SM fields, our model consists of an $SU(2)_L$ singlet vector-like lepton $E$, a second Higgs doublet $H'$ and a dark Higgs field $\phi$, all charged under a $U(1)'$ gauge group, whose gauge boson we denote with $Z'$ \cite{Lee:2020wmh,Lee:2021gnw,Lee:2022nqz}. The $U(1)'$ charge assignments are summarized in Table 1. The charges of the BSM fields are taken to be $+2$ or $-2$ for simplicity, but different choices for the charges can be made rescaling the $g_{_{Z'}}$ coupling. 
	
\begin{table}[hbt!]
	\begin{center}
		\begin{tabular}{c|cccccccccc}
			\hline\hline
			&&&&&&&&&&\\[-2mm]
			& $q_L$ & $u_{R}$  &  $d_{R}$ & $l_{L}$  & $e_{R}$
			& $H$ & $H'$ & $E_L$ & $E_R$ & $\phi$  \\[2mm]
			\hline
			&&&&&&&&&&\\[-2mm]
			$U(1)'$ & $0$ & $0$ & $0$
			& $0$ & $0$ & $0$ & $+2$ & $-2$ & $-2$  & $-2$ \\[2mm]  
			\hline\hline
		\end{tabular}
	\end{center}
	\caption{The $U(1)'$ charges for the SM and extra fields.\label{charges}}
\end{table}	
	
	The Lagrangian for the electroweak sector, including the new fields we introduce, reads
	\begin{align}
	{\cal L}=&-\frac{1}{4}W^a_{\mu\nu}W^{a\, \mu\nu}-\frac{1}{4}B_{\mu\nu}B^{\mu\nu}-\frac{1}{4} F'_{\mu\nu} F^{\prime \mu\nu} - \frac{\sin\xi}{2}  F'_{\mu\nu} B^{\mu\nu}\nonumber\\
	&
	+|D_\mu\phi|^2+|D_\mu H'|^2 +|D_\mu H|^2- V(\phi,H,H')+{\cal L}_{\rm fermions},
	\end{align}
	with
	\begin{eqnarray}
	{\cal L}_{\rm fermions}&=& \sum_{i={\rm SM}, E} i {\bar\psi}_i \gamma^\mu D_\mu \psi_i -y_d {\bar q}_L d_R  H- y_u {\bar q}_L u_R {\tilde H} -y_l {\bar l}_Le_R H \nonumber \\
	&&  -M_E {\bar E}E-\lambda_E \phi {\bar E}_L  e_R-y_E  {\bar l}_L E_R H'  +{\rm h.c.}\,. \label{leptonL}
	\end{eqnarray}
Here, $\tilde H= i \sigma^2 H^*$, $F'_{\mu\nu}$ is the $U(1)'$ field strength, and the covariant derivatives for BSM fields are
\begin{align}
D_\mu\phi&=\left(\partial_\mu +2i g_{_{Z'}} Z'_\mu\right)\phi ,\nonumber\\
D_\mu H'&=\left(\partial_\mu -2i g_{_{Z'}} Z'_\mu-\frac{1}{2}i g_Y  B_\mu- \frac{1}{2} i g \sigma^i W^i_\mu\right) H',\nonumber \\
D_\mu E&=\left(\partial_\mu+2i g_{_{Z'}} Z'_\mu+i g_Y B_\mu\right)E. 
\end{align}
In particular, the dark Higgs $\phi$ is charged only under $U(1)'$, while the vector-like lepton is charged under both $U(1)'$ and $U(1)_Y$, and the second Higgs doublet has the same quantum numbers as the SM Higgs except for its $U(1)'$ charge. 
The potential $V(\phi,H,H')$ is taken as\footnote{Other quartic couplings, $\lambda_5$, $\lambda_6$ and $\lambda_7$, in two Higgs doublet models \cite{2HDM}, are forbidden by the $U(1)'$ symmetry under which the second Higgs doublet $H'$ transforms.}
		\begin{eqnarray}
		V(\phi,H, H') &=& \mu^2_1 H^\dagger H + \mu^2_2 H'^\dagger H'  -(\mu_3 \phi H^\dagger H'+{\rm h.c.})  \nonumber \\
		&&+ \lambda_1 (H^\dagger H)^2 + \lambda_2 (H'^\dagger H')^2+ \lambda_3 (H^\dagger H)(H'^\dagger H')+\lambda_4 (H^\dagger H')(H'^{\dagger}H) \nonumber \\
		&&+ \mu^2_\phi \phi^*\phi + \lambda_\phi (\phi^*\phi)^2+ \lambda_{H\phi}H^\dagger H\phi^*\phi +  \lambda_{H'\phi}H'^\dagger H'\phi^*\phi. \label{scalarpot}
	\end{eqnarray}

In the ``normal" vacuum where the electroweak symmetry and the $U(1)'$ symmetry are broken, we take 
\begin{equation}
	\label{normal vacuum}
	H=\frac{1}{\sqrt 2}\left(\begin{array}{c} 0  \\ v_1 \end{array} \right), \qquad H'=\frac{1}{\sqrt 2}\left(\begin{array}{c} 0  \\ v_2 \end{array} \right), \qquad \phi=\frac{v_\phi}{\sqrt 2}.
\end{equation} 
A mass mixing between the $Z$ and $Z'$ gauge bosons arises in such a vacuum \cite{Lee:2020wmh,Lee:2021gnw,Lee:2022nqz,Bian:2017xzg}. The mass matrix for the charged lepton (muon) and the vector-like lepton \cite{Lee:2022nqz} is defined from 
\bea
{\cal L}_{L,{\rm mass}}&=& -M_E {\bar E}E-m_0 {\bar e}e-( m_R {\bar E}_L e_R+m_L {\bar e}_L E_R+ {\rm h.c.}) \nonumber \\
&=&- ({\bar e}_L, {\bar E}_L) {\cal M}_L  \left(\begin{array}{c} e_R  \\ E_R \end{array} \right) +{\rm h.c.},
 \label{leptonmass0}
\eea
with
\bea
 {\cal M}_L= \left(\begin{array}{cc} m_0 & m_L \\ m_R & M_E \end{array} \right).
\eea
Here, $m_0$ is the bare lepton mass given by $m_0=\frac{1}{\sqrt{2}} y_l v_1$, and $m_R, m_L$ are the mixing masses, given by $m_R=\frac{1}{\sqrt{2}} \lambda_E v_\phi$ and $m_L=\frac{1}{\sqrt{2}} y_E v_2$, respectively.
In the limit of $m_0,m_R, m_L\ll M_E$, the mass eigenvalues for the leptons can be expressed as
\bea
m_{l_2}\simeq M_E, \quad m_{l_1}\approx m_0-\frac{m_R m_L}{M_E}, 
\eea 
while the mixing angles for the right-handed and left-handed leptons  \cite{Lee:2022nqz} become
\bea
\theta_R\simeq  \frac{m_R}{M_E}, \quad \theta_L \simeq  \frac{m_L}{M_E}.
\eea
Therefore, for $m_0\gtrsim\frac{m_R m_L}{M_E} $ we get a seesaw mass contribution for the charged lepton \cite{Lee:2022nqz} as $ \frac{m_R m_L}{M_E}\simeq \theta_R \theta_L M_E$. A similar lepton-portal model with $SU(2)_D$ was considered for muon seesaw mass,  muon $g-2$ as well as $W$ boson mass \cite{Kim:2022zhj}.

\section{Phenomenological constraints on $Z'$}
\label{section: pheno_constraints}
As will be shown in Sec.~\ref{subsec:landau_pole}, in order to explain the muon $g-2$ experimental anomaly with our model while avoiding the appearance of a  Landau pole at too low energies, a small $g_{Z'}$ (correspondingly, a weak-scale $Z'$ boson mass) are necessary.
For such a scenario, the LHC searches for additional $Z'$ gauge bosons should be taken into account.

\subsection{SM anomalies: muon $g-2$ and $W$ boson mass}

The dominant corrections to the muon $g-2$ in our model arise from the one-loop vector-like lepton and $Z'$ gauge boson contributions. In the limit $M_E\gg m_{_{Z'}}$, the correction reads \cite{Lee:2022nqz}
\begin{equation}
\Delta a_\mu\simeq \frac{1}{4\pi^2}\frac{ g^2_{_{Z'}}m_\mu^2}{m^2_{_{Z'}}},
\end{equation}
with $m^2_{_{Z'}}=4g^2_{_{Z'}}(v^2_\phi+v^2\sin^2\beta)$, so that $\Delta a_\mu$ becomes independent of the value of $g_{_{Z'}}$. When $v^2_\phi \gg v^2\sin^2\beta$, $m^2_{_{Z'}}\simeq 4g^2_{_{Z'}}v^2_\phi$ and we only need to fix the VEV of the singlet scalar $v_\phi$ to fit the muon $g-2$.
On the other hand, the correction to the $W$ boson mass in our model stems from the tree-level mixing between the $Z$ and $Z'$ gauge bosons, and for $m_{_{Z'}}\gg m_{_Z} $ it is approximated by
\begin{eqnarray}
\Delta m_{_W}\simeq  \frac{8}{(c^2_{_W}-s^2_{_W})}\,\frac{g^2_{_{Z'}} \, m^2_{_Z}}{g^2_Y\, m^2_{_{Z'}}}\,\,\sin^4\beta.
\end{eqnarray}
From $g^2_{_{Z'}}/m^2_{_{Z'}}\simeq 1/(4v^2_\phi)$ it is immediately understood that, in this limit, the correction to the $W$ boson mass becomes independent of $g_{_{Z'}}$ too.
As $m_{_{Z'}}$ gets closer to $m_{_Z}$, however, the above approximation breaks down, so the smaller $g_{_{Z'}}$ (or $m_{_{Z'}}$), the smaller the correction to $m_{_W}$.

\begin{figure}[!t]
\centering
\includegraphics[width=0.49\textwidth,clip]{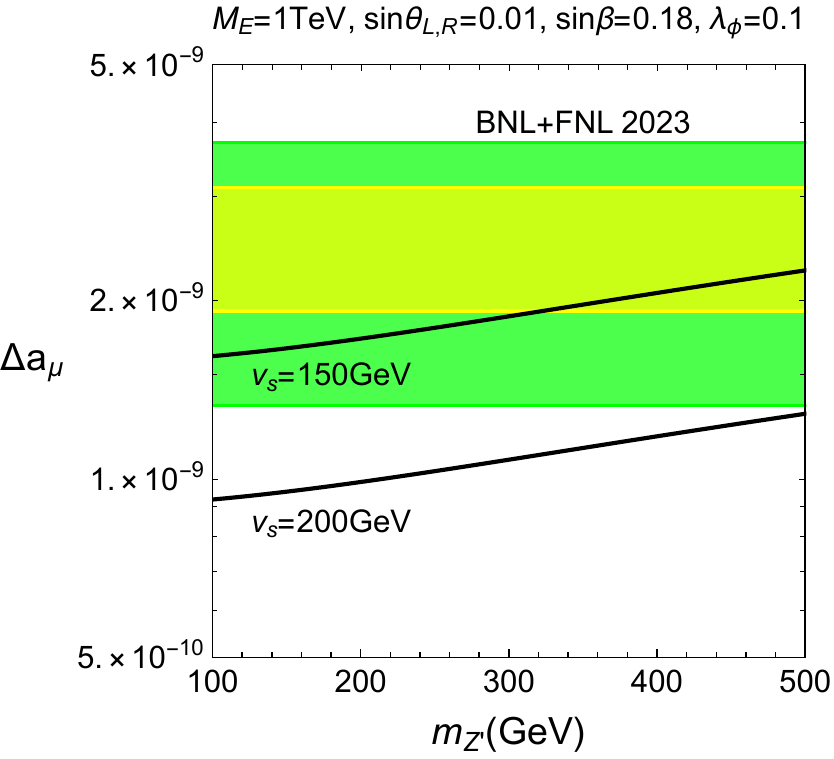}\,\,\,\,\,\,\,\,\,
\includegraphics[width=0.45\textwidth,clip]{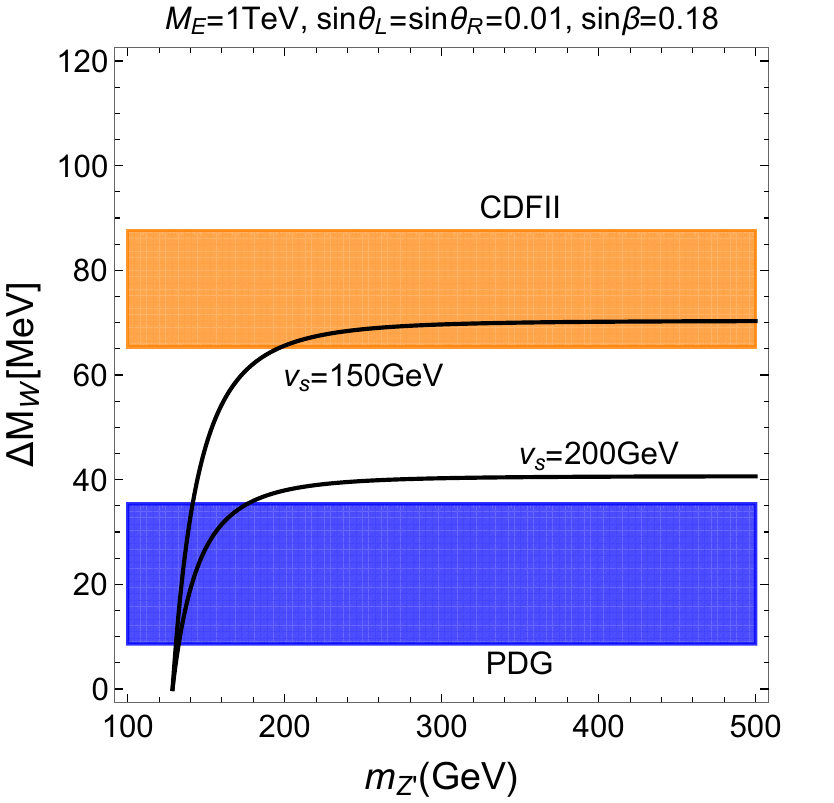} 
\caption{The corrections to the muon $g-2$ (left plot) and the $W$ boson mass (right plot) as a function of $m_{_{Z'}}$. We fixed $M_E=1\,{\rm TeV}$, $\sin\theta_L=\sin\theta_R=0.01$, $\sin\beta=0.18$ and $\lambda_\phi=0.1$. The $1\sigma$ and $2\sigma$ contours of the deviation of the muon $g-2$ from the combined experimental value of BNL and FNL 2023 is shown in yellow and green regions, respectively, and the $1\sigma$ values of the deviation of the $W$ boson mass are shown in orange and blue regions for CDFII and PDG world average, respectively.  
}
\label{anomalies}
\end{figure}

We comment on how flexible the parameter space is to explain the muon $g-2$ and $W$ boson mass anomalies.
In the left plot of Fig.~\ref{anomalies}, we depict the one-loop correction to the muon $g-2$ as a function of $m_{_{Z'}}$ in our model, with $v_s=v_\phi/\sqrt{2}= 200\,{\rm GeV}$ and $150\, {\rm GeV}$. We set the parameters of the vector-like lepton sector to $M_E=1\,{\rm TeV}$, $\sin\theta_L=\sin\theta_R=0.01$, and choose $\sin\beta=0.18$ and $\lambda_\phi=0.1$~\cite{Lee:2022nqz}. In the right plot of Fig.~\ref{anomalies}, we also show the tree-level correction to the $W$ boson mass due to the $Z$-$Z'$ mixing as a function of $m_{_{Z'}}$, with $v_s=v_\phi/\sqrt{2}= 200\,{\rm GeV}$ and $150\, {\rm GeV}$, making the same choice for the other parameters as in the previous case.  We note that in the case of $v_s=200\,{\rm GeV}$, new contributions to the muon $g-2$ in our model are in tension with the experimental value at $2\sigma$ for the SM prediction based on Ref.~\cite{SM}, but they can be consistent with the recent lattice results \cite{lattice2} at $2\sigma$. Thus, the parameter choices for Fig.~\ref{anomalies} are compatible with the SM predictions appearing in the literature, so they will be taken for the later analysis on RG equations.

As a result, we find that a smaller value of $v_s$, for instance, $v_s=150\,{\rm GeV}$, is favoured to explain the muon $g-2$ within $1\sigma$ or $2\sigma$ levels, as shown in the left plot of Fig.~\ref{anomalies}. For the same value of $v_s$, we can explain the $W$ boson mass reported by CDFII for a wide range of $Z'$ boson masses, as shown in the right plot of Fig.~\ref{anomalies}.
For a larger value of $v_s$, for instance $v_s=200\,{\rm GeV}$, we can still explain the $W$ boson mass anomaly from CDFII by taking a mild increase in $\sin\beta$ from 0.18, as far as $m_{_{Z'}}\gtrsim 200\,{\rm GeV}$ \cite{Lee:2022nqz}.

\subsection{$Z'$ signatures at the LHC}

For a relatively light $Z'$ boson, i.e. $m_{_{Z'}}\sim 200-500$~GeV, dilepton final states become crucial for LHC phenomenology. 
We show the $Z'$ production cross section from $pp$ collisions at $13$~TeV and the branching fractions of the $Z'$ boson for the current LHC constraints and prospects. We assume the extra Higgs boson and the dark Higgs boson to have large enough masses so that their production from a $Z'$ decay is kinematically forbidden.
We employ \verb|MadGraph5_aMC@NLO|~\cite{Alwall:2014hca} to calculate the cross sections and branching ratios of the $Z'$, while 
the model is implemented with~\verb|FeynRules|~\cite{Alloul:2013bka}.
\begin{figure}[t!]
	\centering
	\includegraphics[scale=0.5]{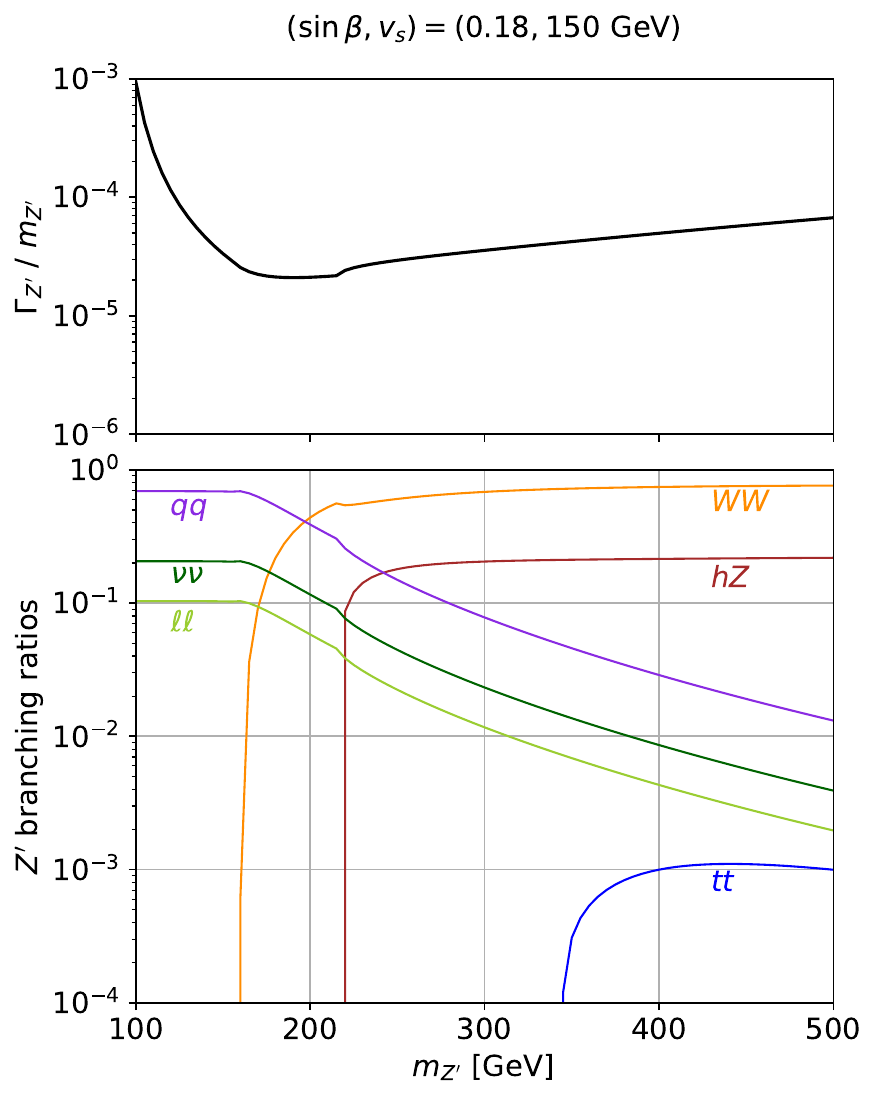}
	\includegraphics[scale=0.5]{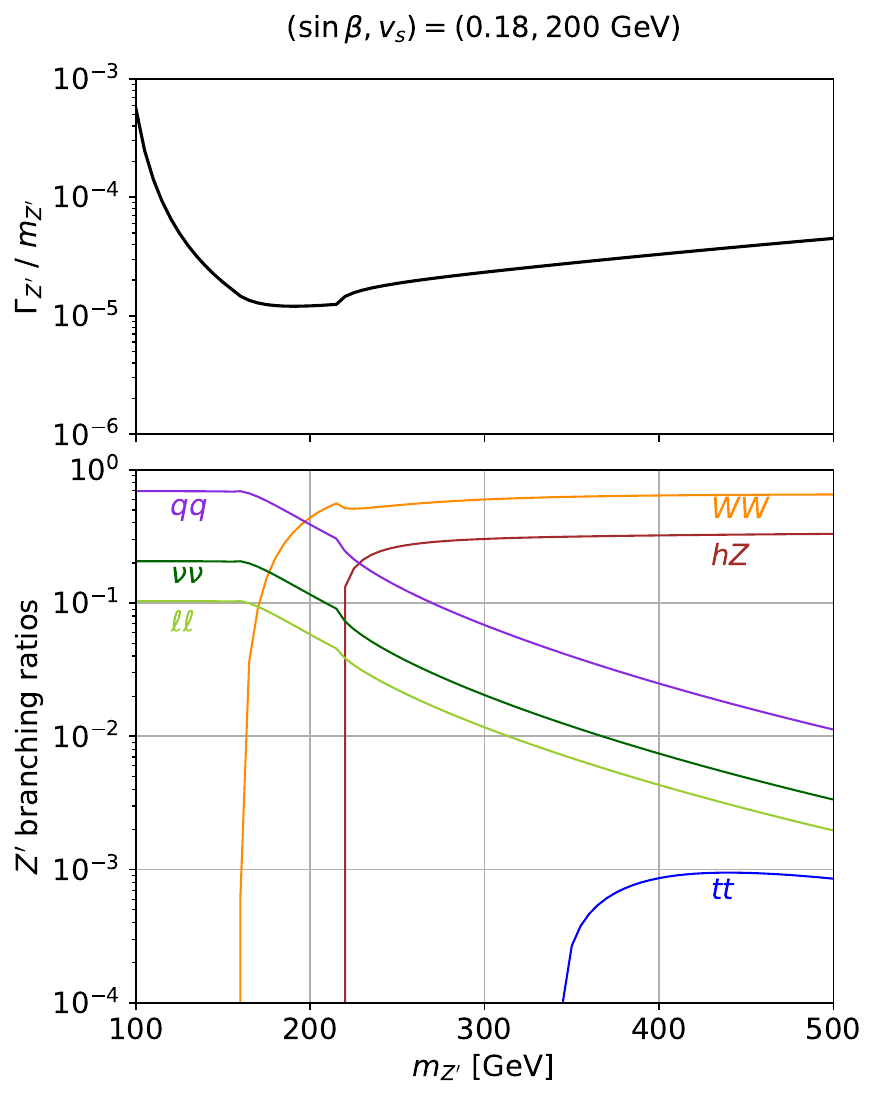}
	\caption{Ratio of $Z'$ total width to its mass and $Z'$ branching ratios as a function of $Z'$ mass, for the dark Higgs VEV of $v_s  = 150$~GeV (left) and  $v_s  = 200$~GeV (right), respectively. We take $\sin\beta = 0.18$; $qq$, $ll$, and $\nu\nu$ include five flavors of quarks, three flavors of leptons, and three flavors of neutrinos, respectively.}
	\label{fig:zp_br}
\end{figure} 

We begin by discussing the $Z'$ boson decays. The $Z-Z'$ mass mixing allows the $Z'$ to decay into states coupled to the $Z$ 
with a suppression factor determined by the $Z-Z'$ mixing angle $\sin\zeta$; the decay width is then suppressed by a factor $\sin^2\zeta$.
When kinematically allowed, then, the $WW$, $hZ$, and $tt$ channels open.
The $U(1)'$ charge of the second Higgs determines an additional interaction for $hZ$ final states, i.e.\,a $Z'$-$Z$-$h$ vertex whose corresponding width is not suppressed by the $Z$ mass mixing angle, but by a $\sin^4\beta$ factor.
The $Z'$-$Z$-$h$ interaction Lagrangian is
\begin{align}
\mathcal{L}_\text{$Z'$-$Z$-$h$} = Z'^{\mu} Z_{\mu} h v\frac{\sec\theta_{_W}}{4}&  \left(\sec\theta_{_W}\sin2 \zeta\left(g^2+4 g_{_{Z'}}^2 \left(\cos2 \beta-2 \sin ^2\beta\cos2\theta_{_W}\right)
-4 g_{_{Z'}}^2 \right) \right. \nonumber\\ &\left. \, \, \, \, \, \, -8 g g_{_{Z'}}  \sin ^2\beta\cos2\zeta\right).
\label{eq:zp-z-h}
\end{align}
Here, $\theta_{_W}$ is the Weinberg angle.
Figure \ref{fig:zp_br} shows the $Z'$ total width scaled by the mass, $\Gamma_{_{Z'}}/m_{_{Z'}}$, and the decay branching ratios for the case where only decays to SM particles are allowed. We take $\sin\beta = 0.18$ and $v_s= 150$~GeV (in the left plot) or $v_s = 200$~GeV (in the right plot) as benchmark points. The $qq$, $ll$, and $\nu\nu$ channels include five flavors of quarks, three flavors of leptons, and three flavors of neutrinos, respectively.
We see that, in addition to the dilepton channel, the $WW$ and $hZ$ channels are important (actually, the dominant ones) when kinematically allowed.
The resonance is always very narrow ($\Gamma_{_{Z'}}/m_{_{Z'}} <  0.1\%$), and, as long as $m_{_{Z'}} \lesssim 2 m_{_{W}}$, $\Gamma_{_{Z'}}/m_{_{Z'}}$  drops as $m_{_{Z'}}$ increases due to the decrease of the $Z-Z'$ mixing.

\begin{figure}[t!]
	\centering
	\includegraphics[scale=0.5]{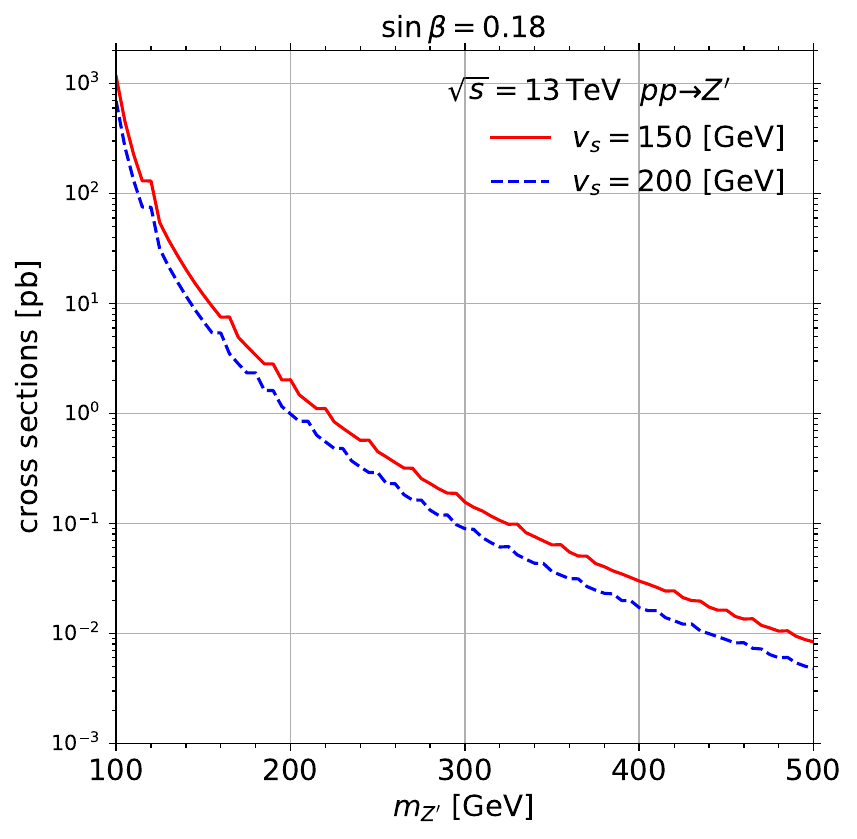}
	\includegraphics[scale=0.5]{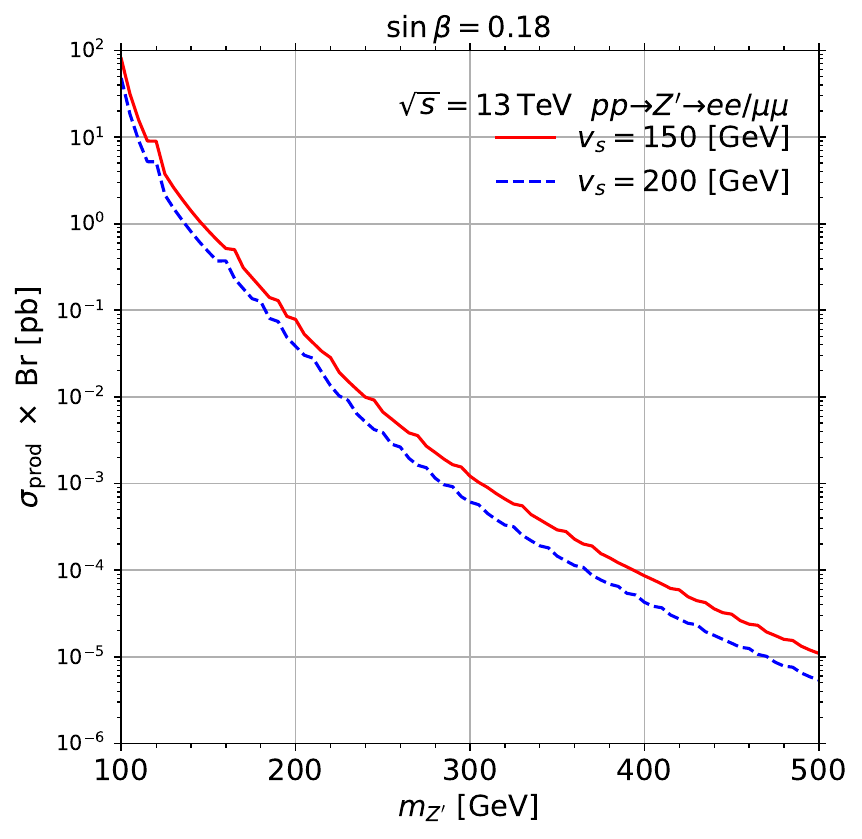}
	\caption{Total cross sections for $Z'$ productions (left) and for $Z'$ productions multiplied by a branching ratio to dilepton ($ee$ and $\mu\mu$) (right)
	at the $13$~TeV LHC as a function of $Z'$ boson mass.
	Two choices of $v_s$ are considered: $150$~GeV shown as a red solid line and $200$~ GeV shown as a blue dashed line.}
	\label{fig:xsec}
\end{figure} 
Let us now discuss the $Z'$ production and its decay to the dilepton state.
The $Z'$ production cross sections (left) and $Z'$ production cross sections multiplied by the $ee$ and $\mu\mu$ dilepton branching ratio (right)
for $pp$ collision at $13$~TeV are depicted in Fig.~\ref{fig:xsec}
as a function of $m_{_Z'}$. 
The opening of the $WW$ and $hZ$ channels is such that the dilepton production is suppressed strongly for large enough values of $m_{_{Z'}}$, for e.g.\,it is less than $1$~fb for $m_{_{Z'}} \gtrsim 300~$GeV.

With the results of ATLAS for dilepton final states at the $13$~TeV LHC run with integrated luminosity $L_\mathrm{int} = 139$/fb~\cite{ATLAS:2019erb},
parametric regions with $m_{_{Z'}} \gtrsim 300~$~GeV and $\sin\beta = 0.18$ are allowed in our scenario.
As we have seen, within this parametric region the $WW$ and $hZ$ final states are dominant, and these resonant boson productions 
are safe with respect to the $L_\mathrm{int} = 139$/fb, $13$~TeV LHC results ~\cite{ATLAS:2022enb,ATLAS:2022qlc,ATLAS:2020pgy,ATLAS:2023alo}.
As the LHC constraints on dileptons are quite severe, dilepton final state produced from $Z'$ boson resonance
around $m_{_{Z'}} \sim 300~$GeV should be targeted by the next LHC results. The same considerations hold for $WW$ final states for $m_{_{Z'}} \gtrsim 300~$ GeV, with the latter being dominant within our scenario.

\section{SM anomalies and running couplings}
\label{Section V1l and RGEs}
	
We begin by discussing some conditions that the scalar potential must satisfy at tree level, and we later consider the one-loop effective potential.
Fixing the low-energy parameters to explain the SM anomalies and taking the alignment limit for the Higgs quartic couplings, we undertake the RG analysis of the model, and in particular of the quartic couplings, and identify the Landau pole and the perturbativity scale.

\subsection{Tree-level stability of the potential}	

The potential $V(\phi,H, H')$ in eq.~(\ref{scalarpot}) has a rich vacuum structure, with CP-breaking and charge-breaking minima, besides CP-even and neutral ones (that we will refer to as ``normal" ones from now on). To derive the conditions under which a normal vacuum can develop, we parametrize the fields in the following way \cite{Kannike:2012pe}: 
\bea
|H|=\chi_1, \quad |H'|=\chi_2,\quad H^\dagger H'=\chi_1 \chi_2 \rho e^{i\theta},\quad \phi=\phi_1 e^{i\varphi}.  
\eea
From the inequality $|H^\dagger H'|\le|H||H'|$, $\rho$ is $\rho\in [-1,1]$. In terms of this new set of parameters, the potential \eqref{scalarpot} takes the form 
\begin{eqnarray}
	\label{potential components l4}
	V(\chi_1,\chi_2,\phi_1,\rho,\theta,\varphi) &=& \mu^2_1 \chi_1^2 + \mu^2_2 \chi_2^2  - 2\mu_3 \cos(\varphi+\theta) \phi_1 \rho \chi_1 \chi_2  + \lambda_1 \chi_1^4 + \lambda_2 \chi_2^4+ \lambda_3 \chi_1^2 \chi_2^2\nonumber \\
	&&+\lambda_4 \chi_1^2 \chi_2^2 \rho^2 
	+ \mu^2_\phi \phi_1^2 + \lambda_\phi \phi_1^4+ \lambda_{H\phi} \chi_1^2\phi_1^2 +  \lambda_{H'\phi}\chi_2^2\phi_1^2,
\end{eqnarray}
and its extremization conditions are then
\begin{equation}
	\label{extremization}
	\begin{cases}
		\left(\mu_1^2+2\lambda_1 \chi_1^2+\lambda_3\chi_2^2+\lambda_4 \rho^2 \chi_2^2+\lambda_{H\phi}\phi_1^2\right)\chi_1-\mu_3\cos\left(\varphi+\theta\right)\rho \phi_1 \chi_2=0,\\
		\left(\mu_2^2+2\lambda_2 \chi_2^2+\lambda_3 \chi_1^2+\lambda_4 \rho^2 \chi_1^2+\lambda_{H'\phi}\phi_1^2\right)\chi_2-\mu_3\cos\left(\varphi+\theta\right)\rho\phi_1 \chi_1=0, \\
		\left(\mu_\phi^2+2\lambda_\phi \phi_1^2+\lambda_{H\phi}\chi_1^2+\lambda_{H'\phi}\chi_2^2\right)\phi_1-\mu_3\cos\left(\varphi+\theta\right)\rho \chi_1 \chi_2=0,\\
		\lambda_4 \chi_1^2 \chi_2^2\rho-\mu_3 \cos\left(\varphi+\theta\right)\phi_1 \chi_1 \chi_2=0,\\
		\sin\left(\varphi+\theta\right)\phi_1 \rho \chi_1 \chi_2=0. 
	\end{cases}
\end{equation} 
When $\phi_1, \rho, \chi_1, \chi_2\ne 0$, the last equation is solved by $\sin\left(\varphi+\theta\right)=0$, from which $\cos\left(\varphi+\theta\right)=\pm 1$. When $\mu_3 \phi_1\rho \chi_1 \chi_2> 0$, the upper sign minimizes the potential and the lower sign maximizes it. The opposite is true for $\mu_3 \phi_1\rho \chi_1 \chi_2<0$. 

Let us now concentrate on the value of $\rho$ that minimizes the potential. We start with the case $\mu_3 \phi_1\rho \chi_1 \chi_2> 0$, and thus $\cos\left(\varphi+\theta\right)=1$. For a positive $\lambda_4$, the $\lambda_4 \chi_1^2 \chi_2^2 \rho^2$ contribution in the potential is necessarily positive and contrasts the  negative contribution given by $\mu_3$. The value of $\rho$ at the minimum is given by the solution to eq.\,$\eqref{extremization}_4$ (keeping in mind that $|\rho|\le 1$). On the contrary, when $\lambda_4<0$, the $\lambda_4 h_1^2 h_2^2 \rho^2$ contribution is necessarily negative. It adds up with the negative contribution coming from the $\mu_3$ term. In this case, the minimum is found for $\rho=\pm 1$, and the solution to $\eqref{extremization}_4$ maximizes the potential. 
When $\mu_3 \phi_1\rho \chi_1 \chi_2<0$, we should take $\cos\left(\varphi+\theta\right)=-1$. As a consequence, the contribution from $\mu_3$ is again negative, and the same observations as those made in the $\mu_3 \phi_1\rho \chi_1 \chi_2>0$ case apply. 

The normal vacuum in eq.~\eqref{normal vacuum} corresponds to $h_1^2=v_1^2/2$, $h_2^2=v_2^2/2$, $\rho=1$, $\theta=2\pi n$, $\phi_1^2=v_\phi^2/2$ , $\varphi=2\pi n$. In particular, we have $h_1, h_2, \phi_1, \rho>0$ and $\cos\left(\varphi+\theta\right)=1$. 
Then, from the above discussion, it is immediate to deduce that there are two scenarios where the configuration for the normal vacuum in eq.~\eqref{normal vacuum} can appear as a minimum of the potential: when $\mu_3>0$ and $\lambda_4<0$, or when $\mu_3>0$, $\lambda_4>0$ and the solution to $\eqref{extremization}_4$ gives $\rho=1$. For this to happen, $\lambda_4$ must be $\lambda_4=\sqrt2 \mu_3 v_\phi/v_1 v_2$. As we will see in the following, the values of $\mu_3$, $v_1$, $v_2$ and $v_\phi$ that we will be concerned with in our analysis are such that the vacuum solution with positive $\lambda_4$ is not of interest, as it would require too large values of $\lambda_4$.

Our analysis will only be concerned with normal vacua, and, as usually done, we will parametrize the VEV of the two Higgs doublets as
$v_1= v\cos\beta$, $v_2=v\sin\beta$, where $v$ is the value of the Higgs VEV in the Standard Model.
Concerning the other types of vacua mentioned above, let us remind that, as shown in \cite{Ferreira:2004yd,Barroso:2005sm,Ivanov:2006yq,Ivanov:2007de}, minima of a different nature cannot simultaneously coexist at tree-level in the 2HDM. Besides the impossibility of having minima of different nature at tree-level, it is well known that multiple non-degenerate vacua of the same nature can simultaneously coexist at tree-level in the 2HDM \cite{Barroso:2007rr}. The parameter space must then be restricted to those regions where either the potential does not develop any minimum deeper than the electroweak-like one or, if  it does develop such a minimum, the tunneling time from the electroweak-like vacuum is larger than the age of the Universe, $\tau\ge \tau_{_U}$ \cite{Branchina:2018qlf}. The benchmark points that will be considered in our analysis do not develop any minimum different than the SM one at tree-level, and we do not need to worry about these issues. Stability of the potential will be investigated at the radiative (one-loop) level in the later discussion.

Tree-level stability of the potential requires that the quartic coupling matrix\footnote{In the following equations, we keep a generic $\rho$. For the normal background configurations considered in this work, $\rho$ is $\rho=1$.} (in the basis $\left(\chi_1^2, \chi_2^2,\phi_1^2 \right)$),
\begin{align}
	\mathcal M_\lambda =\left(\begin{array}{ccc} \lambda_1 &  \frac{\lambda_3+\rho^2\lambda_4}{2} & \frac{\lambda_{H\phi}}{2} \vspace{0.2cm} \\ \frac{\lambda_3+\rho^2\lambda_4}{2} & \lambda_2 & \frac{\lambda_{H'\phi}}{2} \vspace{0.2cm}   \\ \frac{\lambda_{H\phi}}{2} & \frac{\lambda_{H'\phi}}{2}  &  \lambda_\phi \end{array} \right),
\end{align}
is copositive definite \cite{Kannike:2012pe}. This gives the conditions
\begin{equation}
	\begin{cases}
		\lambda_1\ge 0,\qquad \lambda_2\ge0,\qquad \lambda_\phi\ge0, \\
		\lambda_3+\rho^2\lambda_4+2\sqrt{\lambda_1\lambda_2}\ge0,\qquad \lambda_{H\phi}+2\sqrt{\lambda_1\lambda_\phi}\ge 0,\qquad \lambda_{H'\phi}+2\sqrt{\lambda_2\lambda_\phi}\ge 0,
	\end{cases}
	\label{conditions stability}
\end{equation}  
and 
\begin{align}
	&\sqrt{\lambda_1 \lambda_2\lambda_\phi}+\frac{\lambda_3+\rho^2\lambda_4}{2}\sqrt{\lambda_\phi}+\frac{\lambda_{H\phi}}{2}\sqrt{\lambda_2}+\frac{\lambda_{H'\phi}}{2}\sqrt{\lambda_1}\nonumber\\
	&+\frac{1}{2}\sqrt{(\lambda_3+2\sqrt{\lambda_1\lambda_2})(\lambda_{H\phi}+2\sqrt{\lambda_1\lambda_\phi})(\lambda_{H'\phi}+2\sqrt{\lambda_2\lambda_\phi})}\ge 0.
	\label{condition stability 2} 
\end{align}	
Having in mind the typical RG improvement of the potential, in the remainder we will study radiative stability of the potential at the one-loop level by requiring that the above conditions \eqref{conditions stability}, \eqref{condition stability 2} are satisfied with the running couplings $\lambda_i(\mu)$.

\subsection{One-loop effective potential}	
	
The gauge structure of the model is such that the scalar degrees of freedom can be restricted from $10$ to $6$. As we are only interested in a specific type of configurations and want to study the potential only in the subspace of neutral CP-even scalars, we further restrict the problem to a $3$-field problem. 
	
We parametrize the doublets and the scalar singlet as
		\begin{eqnarray}
			\label{parametrization H1}
	H&=&\left(\begin{array}{cc} \phi^+_1 \\ \frac{1}{\sqrt{2}}(h_1+\rho_1+i\eta_1) \end{array} \right), \\
		\label{parametrization H2}
	H'&=&\left(\begin{array}{cc} \phi^+_2 \\ \frac{1}{\sqrt{2}}(h_2+\rho_2+i\eta_2) \end{array} \right),\\
		\label{parametrization phi}
	\phi&=&\frac{1}{\sqrt{2}}(s+\rho_3+i\eta_3)
	\end{eqnarray}
where $h_1$, $h_2$, $s$ are the background fields of which we want to study the potential\footnote{As discussed in the Appendix, to distinguish contributions that renormalize the $|H|^2|H'|^2$ operator from contributions that renormalize the $|H^\dagger H'|^2$ operator, we need to consider a slightly more complicated background configuration, where at least one of the doublets develop an upper component. The $(h_1,h_2,s)$ background of equations \eqref{parametrization H1}-\eqref{parametrization phi} can be re-obtained taking the limit for the upper component to vanish after renormalization is performed and renormalization group equations have been determined. }. The mass matrices obtained in the $(h_1, h_2, s)$ background (actually, in a slightly more general background) are easily calculated and are reported in Appendix \ref{Appendix: A}.
	
The tree-level potential for $h_1, h_2$ and $s$ is obtained straightforwardly from the expansion of \eqref{scalarpot},
\begin{align}
V_{0}(h_1,h_2,s)&=\frac{\mu_1^2}{2}h_1^2+\frac{\mu_2^2}{2}h_2^2+\frac{\mu_\phi^2}{2}s^2+\frac{\lambda_1}{4}h_1^4+\frac{\lambda_2}{4}h_2^4+\frac{\lambda_\phi}{4}s^4\nonumber\\
&+\frac{\lambda_3+\lambda_4}{4}h_1^2h_2^2+\frac{\lambda_{H\phi}}{4}h_1^2s^2+\frac{\lambda_{H'\phi}}{4}h_2^2s^2-\frac{\mu_3}{\sqrt 2}h_1h_2s.
\label{V0}
\end{align}
Performing the typical loop integral with the mass terms $\mathcal M_i(h_1,h_2,s)$ and indicating with $\Lambda$ the cutoff of the theory, we obtain in the physically meaningful limit $\Lambda^2\gg M_i^2$  ($M_i$ indicate the eigenvalues of the mass matrices)
\begin{equation}
V_{1l}=\sum_i (-1)^{\delta f,i} n_i \frac{{\rm Tr}\left( \mathcal M_i^2\right)}{32\pi^2}\Lambda^2+\sum_i (-1)^{\delta f,i} \frac{n_i}{64\pi^2}{\rm Tr} \mathcal M_i^4\left(\log\left(\frac{\mathcal M_i^2}{\Lambda^2}\right)-\frac{1}{2}\right)+\mathcal O(\Lambda^{-1}),
\label{V1}
\end{equation}
where $n_i=12$ for the top quark, $n_i=4$ for the VLL-muon fermion mass matrix \eqref{fermion mass matrix}, $n_i=3$ for the gauge mass matrix \eqref{gauge mass matrix} and $n_i=1$ for the scalar one \eqref{scalar mass matrix}.
If we write the quadratic contribution in $\Lambda$ as $\frac{1}{2}\left(\alpha_1 h_1^2+\alpha_2 h_2^2+\alpha_\phi s^2\right)\Lambda^2$, we have
\begin{equation}
	\label{quad UV-sensitive}
	\begin{cases}
		\alpha_1=\frac{9 g^2}{64\pi^2}+\frac{3g_Y^2}{64 \pi^2}+\frac{3\lambda_1}{8 \pi ^2}+\frac{\lambda_3}{8 \pi^2}+\frac{\lambda_4}{16\pi^2}+\frac{\lambda_{H\phi}}{16 \pi^2}-\frac{y_l^2}{8 \pi ^2}-\frac{3 y_t^2}{8 \pi^2}, \\		
		\alpha_2=\frac{9 g^2}{64 \pi^2}+\frac{3 g_Y^2}{64 \pi^2}+\frac{3g_{_{Z'}}^2}{4 \pi^2}+\frac{3\lambda_2}{8 \pi ^2}+\frac{\lambda_3}{8 \pi^2}+\frac{\lambda_4}{16\pi^2}+\frac{\lambda_{H'\phi}}{16 \pi^2}-\frac{y_E^2}{8 \pi ^2},\\
		\alpha_3=\frac{3 g_{_{Z'}}^2}{4 \pi ^2}+\frac{\lambda_\phi}{4 \pi ^2}+\frac{\lambda_{H\phi}}{8 \pi ^2}+\frac{\lambda_{H'\phi}}{8 \pi ^2}-\frac{\lambda_E^2}{8 \pi ^2}.  
	\end{cases}
\end{equation}	
In the above equations we kept the contributions from the muon Yukawa coupling $y_l$. The reason we keep terms proportional to $y_l$ in our calculation is that the latter enters in the mass matrix that describes the mixing between the muon and the vector-like lepton. As such, it might happen that $y_l$ appears in combination with other couplings that enhance its contribution. 

The Renormalization Group Equations (RGEs) are conveniently extracted requiring independence from $\Lambda$ of the full effective potential at the one-loop level, $V=V_0+V_{1l}$, namely,
\begin{equation}
	\label{RGE potential}
\Lambda\frac{d}{d\Lambda}V=0\rightarrow \left(\Lambda\frac{\partial}{\partial\Lambda}+\mu_i^2\gamma_{\mu_i^2}\frac{\partial}{\partial\mu_i^2}+\beta_{\mu_3}\frac{\partial}{\partial\mu_3}+\beta_{\lambda_i}\frac{\partial}{\partial\lambda_i}-\gamma_i\phi_i\frac{\partial}{\partial\phi_i}\right)V=0
\end{equation}
where $i$ runs over $h_1,h_2,s$ while $\beta$ and $\gamma$ indicate the beta functions and anomalous dimensions, respectively. We show the various RGEs extracted from this equation, together with all the other necessary RGEs, in Appendix \ref{Appendix: B}. 

To implement the fact that  the contribution from a state of running mass $m(\mu)$ is approximately frozen at scales $\mu$ below $m(\mu)$ \cite{Branchina:2022gll},  we supplement the RGEs found from the one-loop calculations, and presented in Appendix \ref{Appendix: B}, with threshold corrections in the form of the Heaviside theta function $\theta(\mu^2-m^2(\mu))$.

\subsection{Conditions from SM anomalies}

Taking into account the parameter space motivated by both the muon $g-2$ and the $W$ boson mass in Ref.~\cite{Lee:2022nqz}, we consider the following set of boundary values at low energy\footnote{As compared to Ref.~\cite{Lee:2022nqz}, we use the convention $\phi=(s+i\eta_3)/\sqrt 2$. Accordingly, the vacuum value of $\phi$, $v_\phi$, in our work, corresponds to $\sqrt 2$ times the vacuum value of $\phi$ in \cite{Lee:2022nqz}.}:
\begin{equation*}
\begin{cases}
M_E=1\, {\rm TeV}, \\
\sin\beta=0.18, \\
\sin\theta_L=0.01 \rightarrow y_E= 0.32, \\
\sin\theta_R=0.011 \rightarrow \begin{cases}
	\text{if } v_\phi= \sqrt{2}\times 200\, {\rm GeV} \rightarrow \lambda_E= 0.055\,;\\
	\text{if } v_\phi= \sqrt{2}\times 150\, {\rm GeV} \rightarrow \lambda_E= 0.073,
\end{cases}\\
m_{_{Z'}}=500\, {\rm GeV} \rightarrow \begin{cases}
	\text{if }v_\phi= \sqrt{2}\times 200\, {\rm GeV} \rightarrow  g_{_{Z'}}= 0.87\,;\\
	\text{if }v_\phi= \sqrt{2}\times 150\, {\rm GeV} \rightarrow  g_{_{Z'}}= 1.15.
\end{cases}
\end{cases}
\end{equation*}  
Here, we took two cases for the VEV of the singlet scalar, $v_s=v_\phi/\sqrt{2}= 200\,{\rm GeV}$ and $v_s=150\, {\rm GeV}$.
From $m^2_{_{Z'}}=4 g_{_{Z'}}^2\left(v^2\sin^2\beta+v_\phi^2\right)$, which is the diagonal term of the neutral gauge boson mass matrix $\mathcal M^2_Z$ in the $Z'$ direction, we can determine $g_{_{Z'}}$ for given $\sin\beta$ and $m_{_{Z'}}$. Moreover, we can also determine $\lambda_E$ for seesaw muon mass.

We now list up the procedure for determining the boundary values of the rest of the parameters in our model.
From $m_{_W}=gv/2$, the IR boundary condition for $g$ is the same as in the SM. Concerning the $Z$ boson, $m_{_Z}=\sqrt{g^2+g_Y^2}\, v/2$ is the diagonal term of the $\mathcal M^2_Z$ matrix in the $Z$ direction. The boundary value for $g_Y$ is found imposing that the smaller eigenvalue of  $\mathcal M^2_Z$ is the measured $Z$ mass. Finally, since the top and the muon only couple to $h_1$, but not to $h_2$, the bare mass terms generated by the scalar VEVs for them are $m_t=y_t v_1/\sqrt2$ and $m_l=y_l v_1/\sqrt 2$, respectively. This means that the boundary conditions for $y_t$ and $y_l$ are not the same as in the SM, but are rather magnified by a factor $\cos^{-1}\beta$.We note that the seesaw mass generated by the mixing with the vector-like lepton gives a negative contribution to the muon mass\footnote{In the seesaw limit, the lowest eigenvalue of the fermion mass matrix is $m_{l_1}\sim m_0-\frac{m_R m_L}{M_E}$. The symbols $m_0,\, m_R$ and $m_L$ are defined in Appendix A.} \cite{Lee:2022nqz}, so $y_l$ can be larger than the one in the SM. To avoid a significant tuning for the muon mass, we keep $y_l$ of order similar to the SM one, namely $y_l=y_l^{({\rm SM})}\cos^{-1}\beta$.
 
After imposing the above phenomenological constraints, we still have the remaining $11$ parameters to fix: $\mu_1,\mu_2,\mu_\phi, \lambda_1,\lambda_2,\lambda_\phi,\lambda_3,\lambda_4,\lambda_{H\phi},\lambda_{H'\phi},\mu_3$. To proceed further, we continue to follow \cite{Lee:2022nqz} and take the following steps. 

Some of the non-diagonal terms of the scalar mass matrix $\mathcal M^2_H$ (see Eq.\,\eqref{scalar mass matrix} and the following discussion),  namely, $M^2_{H, 13}$ and $M^2_{H, 23}$,  can be put to zero in the vacuum $(v_1,v_2,v_\phi)$ by the convenient choice,
\begin{equation}
	\label{decoupling darkH}
\lambda_{H\phi}=\frac{\mu_3 \tan\beta}{\sqrt2 v_\phi}, \qquad \lambda_{H'\phi}=\frac{\mu_3 \cot\beta}{\sqrt2 v_\phi}.
\end{equation}
These conditions decouple the dark Higgs $\phi$ from the two Higgs doublets $H$ and $H'$.
With this choice, the third minimization condition in \eqref{extremization} reads $\mu_\phi^2+\lambda_\phi v_\phi^2=0$, and $M^2_{H, 33}$ gives rise to the mass eigenvalue for the singlet scalar,
\begin{equation}
m^2_\phi\equiv M^2_{H, 33}=2\lambda_\phi v_\phi^2+\frac{\mu_3}{2\sqrt 2}\sin(2\beta) \frac{v^2}{v_\phi}.
\end{equation} 

The $2\times 2$ upper matrix of $\mathcal M^2_H$, that we denote by $\widetilde{\mathcal M}^2_H$, can be simplified using the minimization equations in \eqref{extremization}. It takes the form 
\begin{eqnarray}
	\label{Mtilda}
	\widetilde {\mathcal M}^2_H =\left(
	\begin{array}{cc} 2\lambda_1 v_1^2 +\frac{\mu_3}{\sqrt 2}\tan\beta \,v_\phi &  \lambda_{34} v_1 v_2-\frac{\mu_3}{\sqrt 2}v_\phi \vspace{0.2cm} \\ \lambda_{34} v_1 v_2-\frac{\mu_3}{\sqrt 2}v_\phi & 2\lambda_2 v_2^2 +\frac{\mu_3}{\sqrt 2}\cot\beta \,v_\phi  \vspace{0.2cm} 
	\end{array} \right),
\end{eqnarray}
where we defined $\lambda_{34}\equiv \lambda_3+\lambda_4$. 
Taking the $\tan\beta$-independent solution for the alignment limit of the two doublets as in \cite{Lee:2022nqz}, that amounts to set the relations 
\begin{eqnarray}
\lambda_1=\lambda_2=\frac{\lambda_{34}}{2},
\label{aligned}
\end{eqnarray}
among the quartic couplings, a further simplification arises.   
With this input, in fact, we get the eigenvalues of $\widetilde {\mathcal M}^2_H$ as\footnote{For simplicity, below we use the same symbols as for the interaction eigenstate and indicate the mass eigenstates with $h_1$ and $h_2$.}
\begin{equation}
	\label{eigenvalues aligned case}
m^2_{h_1}=2\lambda_1 v^2, \qquad m^2_{h_2}=\sqrt 2 \mu_3 v_\phi \sin^{-1}(2\beta).
\end{equation}
To recover the SM Higgs in the decoupling limit of $H'$ and $\phi$, we identify the first of these two physical masses with the SM Higgs mass. Therefore, the boundary value of $\lambda_1$ (and consequently that of $\lambda_2$ and $\lambda_{34}/2$) is the same one as the boundary value for the Higgs quartic coupling in the SM. As first shown in \cite{Pilaftsis:2016erj} and further stressed in \cite{Lee:2022nqz}, it is interesting to observe that, when the above condition \eqref{aligned} is satisfied, the alignment of the two Higgs doublets is motivated by the fact that the $H$-$H'$ scale invariant sector of the scalar potential is maximally symmetric.

The combination of all the relations and conditions described above leaves us with only three free independent parameters. We can conveniently choose them to be $\lambda_\phi$, $\mu_3$ and $\lambda_3$. In fact, from $\mu_\phi^2+\lambda_\phi v_\phi^2=0$ we see that the choice of $\lambda_\phi$ determines the value of $\mu_\phi$. The choice of $\mu_3$ determines both $\lambda_{H\phi}$ and $\lambda_{H'\phi}$ from \eqref{decoupling darkH}, and $\mu_1$ and $\mu_2$ from the minimization equations \eqref{extremization}$_1$ and \eqref{extremization}$_2$. Finally, the values of $\lambda_3$ and $\lambda_4$ are subject to the condition $\lambda_{34}=2\lambda_1$, so that the choice of one fixes the other. 
As the dependence of the results on the choice of the couple $(\lambda_3,\lambda_4)$ that realizes the condition $\lambda_{34}=2\lambda_1$ is absolutely straightforward, in our investigation we will study the $(\lambda_\phi,\mu_3)$ parameter space and, for each of the points in it, we will consider the choice $(\lambda_3,\lambda_4)=(2\lambda_1,0)$. In this respect, we recall that, for consistency, $\lambda_4$ should be $\lambda_4\le 0$, so any other choice would have to be of the form $(\lambda_3,\lambda_4)=(2\lambda_1+\delta, -\delta)$, with $\delta>0$. It is easily seen that any admissible choice, that is any choice that keeps the couplings in the perturbative regime at sufficiently low energies, gives, for all practical reasons, the same results.

\subsection{Running couplings from the alignment limit in the IR}
 
Having established the number of independent parameters of the presently considered scenario in the previous subsection, we now consider some further constraints on the $(\lambda_\phi, \mu_3)$ parameter space and perform the RG analysis of the running couplings.

A necessary requirement for a BSM theory to be phenomenologically viable is that new particle states are heavier than the SM ones if they are strongly coupled to the SM.  In the scalar sector of our model, and within the scenario considered thus far, $h_2$ and $s$ are heavier than the Higgs provided that
\begin{equation}
	\label{phys mass bounds}
	\begin{cases}
\mu_3>\sqrt{2} \lambda_1 \sin(2\beta)\frac{v^2}{v_\phi}, \vspace{0.2cm} \\
2\lambda_\phi v_\phi^2 +\frac{\mu_3}{2\sqrt 2} \sin(2\beta)\frac{v^2}{v_\phi}> 2\lambda_1 v^2,
\end{cases}
\end{equation}
respectively.
On top of that, perturbativity requires, among others, $\lambda_\phi,\lambda_{H\phi},\lambda_{H'\phi}\le 4\pi$ and $\mu_3^2\lesssim 4\pi \bar m^2$, where $\bar m$ is the heaviest among the physical masses of the scalars $h_1$, $h_2$, and $s$.

A further restriction on the $(\lambda_\phi, \mu_3)$ parameter space arises from the stability conditions  \eqref{conditions stability} of the potential. In fact, from the first line in \eqref{conditions stability}, $\lambda_\phi$ must be $\lambda_\phi\ge 0$. The other inequalities in \eqref{conditions stability}, \eqref{condition stability 2} are then all trivially respected in this scenario.

\begin{figure}[t]
	\centering
	\includegraphics[scale=0.58]{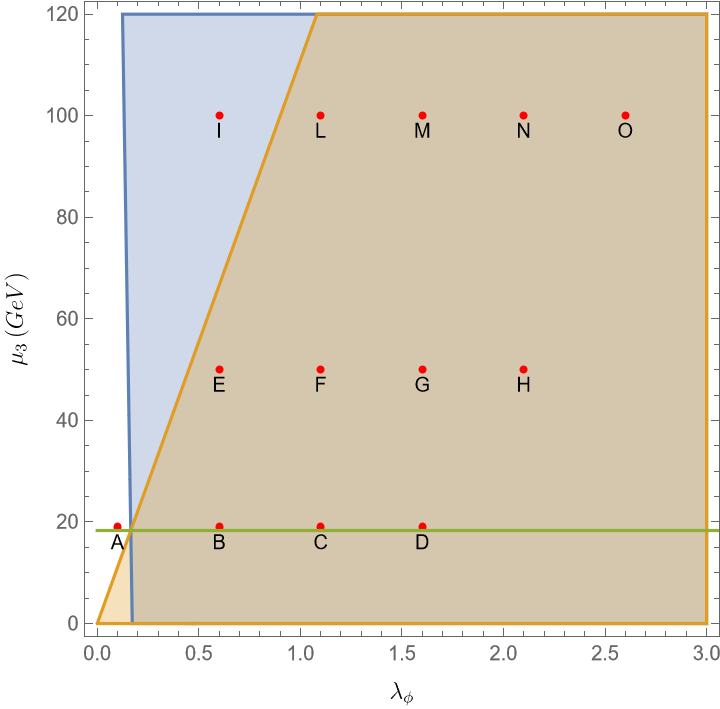}\,\,\,
	\includegraphics[scale=0.58]{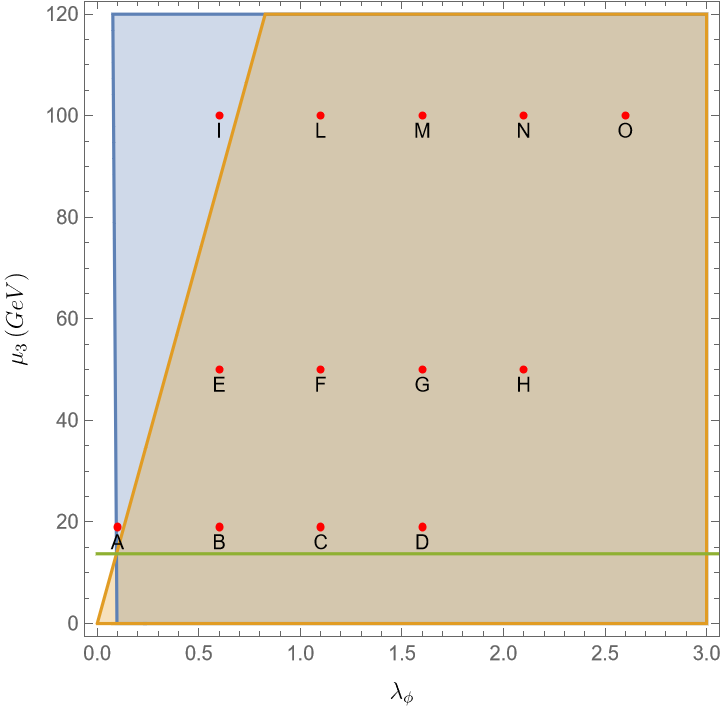}
	\caption{Benchmark points in the parameter space for $\lambda_\phi$ vs $\mu_3$. We chose  $v_\phi=\sqrt2\times150$ GeV on {\it left panel} and $v_\phi=\sqrt2\times200$ GeV on {\it right panel}. 
	In both panels, the blue region indicates the portion of parameter space where the second condition in \eqref{phys mass bounds} ($m_s\ge m_{h_1}$) is respected. Above the light green curve, the first condition in \eqref{phys mass bounds} ($m_{h_2}>m_{h_1}$) is respected. We have $m_s>m_{h_2}$ in the orange region. The points are benchmark points to be presented in Table 2 and Table 3.}
	\label{exclusion plot}
\end{figure}

The relevant bounds and the hierarchies between physical masses are shown in Fig. \ref{exclusion plot}. In the blue and orange regions,  the first condition in \eqref{phys mass bounds} ($m_{h_2}>m_{h_1}$) and $m_s>m_{h_2}$ are  satisfied, respectively. In the region above the light green curve,   the first condition in \eqref{phys mass bounds} ($m_{h_2}>m_{h_1}$) is respected. However, $m_s>m_{h_1}$ and $m_s>m_{h_2}$ are not necessary conditions when the dark Higgs is decoupled from the two doublets. They are rather useful to show the hierarchy between the masses, from which we deduce the relevant unitarity condition as $\mu_3^2\lesssim 4\pi \bar m^2$ where $\bar m^2$ is the heaviest of the three masses. On the contrary, $m_{h_2}>m_{h_1}$ is necessary to get the Higgs couplings consistent with the LHC results and avoid the current LHC bounds on extra scalars in 2HDMs.  We indicated with a red dot the benchmark points that will be taken later to analyse the running of the parameters in Table 2 and Table 3.

Performing a numerical investigation of the parameter space, we find as a a general feature that the couplings tend to develop a singularity at unacceptably low scales (perturbativity is then clearly lost at lower energies). It is easily understood, comparing to analytic solutions of isolated RGEs for a single coupling, that the singularity encountered by the numerical algorithm is nothing but the Landau pole of the theory (see below for an example of such a comparison). The reason for such low Landau poles is explained below.

Except for $\lambda_1$, the beta functions of the scalar couplings in our model are strongly unbalanced in favour of the bosonic contributions. 
The IR boundary values of the additional fermion couplings are too small to counterbalance them. For instance, the Yukawa couplings are taken as $y_E= 0.32$ while the IR value of $\lambda_E$ varies from $\lambda_E=0.073$ to $\lambda_E=0.055$ for $v_s= 150$ GeV and $v_s=200$ GeV, respectively. The IR value of the $Z'$ gauge coupling takes the value $g_{_{Z'}}= 1.15$ and $g_{_{Z'}}=0.87$  for $v_s= 150$ GeV and $v_s= 200$ GeV, so it is largely the dominant one in the RG equations and rapidly drives the scalar couplings to large values. On top of that, the greater the value of $\mu_3$, the greater the IR values of $\lambda_{H'\phi}$ and $\lambda_{H\phi}$, so that positive bosonic contributions are even more unbalanced as we take larger values of the cubic coupling $\mu_3$. In this respect, it is worth to note that the decoupling condition \eqref{decoupling darkH} is such that $\lambda_{H'\phi}\sim 30 \lambda_{H\phi}$. For the lowest values of $\mu_3$ in agreement with \eqref{phys mass bounds} and $\lambda_\phi=0.1$ (phenomenologically we need even larger values of $\mu_3$ for the additional scalars to be heavy enough), we find a singularity around $\mu= 31$ TeV for $v_s=150$ GeV and around $\mu= 636$ TeV for $v_s=200$ GeV.

Table $2$ and Table $3$ below show the scales where the Landau pole appears and  perturbativity is lost for the benchmark points in the left and right plots of Fig.~\ref{exclusion plot}, respectively. The way these scales vary in the $(\lambda_\phi, \mu_3)$ parameter space is easily inferred from the results. As expected, the smaller the values of $\lambda_\phi$ and $\mu_3$, the higher these scales are. However, they do not show a very large sensitivity to changes in the values of $\lambda_\phi$ and $\mu_3$. As already mentioned, the sensitivity to the specific choice of the $(\lambda_3,\lambda_4)$ couple is much more suppressed, and is typically beyond the accuracy of our investigation. Nevertheless, it can be seen that among all the possible choices that respect the condition $\lambda_3+\lambda_4=2\lambda_1$ with $\lambda_4\le0$, the one we chose, that is $\lambda_3=2\lambda_1,\, \lambda_4=0$, is the one with the highest Landau pole. 

The two cases considered $v_s= 150$ GeV and $v_s=200$ GeV correspond to the central value and the highest possible value found in \cite{Lee:2022nqz} to fit the muon $g-2$ within $2\sigma$, respectively, but they predict dangerously low scales for Landau pole and perturbativity. 
As can be easily inferred from a comparison of Table \ref{table:2} and Table \ref{table:3}, models with larger values of $v_s$ push the problematic scales to slightly higher values. This is mainly due to the fact that $g_{_{Z'}}$ is inversely proportional to $v_s$. In fact, we will see in next section that the $\sim $ TeV scale Landau pole is mainly generated by the too large IR value of $g_{_{Z'}}$ considered in \cite{Lee:2022nqz}.

\begin{table}[htbp!]
	\centering  
	\begin{tabular}{|p{1.2cm}|p{1.1cm}|p{1.1cm}|p{2.7cm}|p{2.5cm}|p{1.1cm}|p{1.1cm}|p{1.1cm}|p{1.1cm}|}
		\hline
		\multicolumn{9}{|c|}{Benchmark points in Fig.\,\ref{exclusion plot} (Left): $v_\phi=\sqrt{2}\times 150$ GeV} \\
		\hline Point & 
		$\lambda_\phi$ & $\mu_3$ (GeV) & Landau pole\,($\times10^4\,$GeV) & Perturbativity ($\times10^3\,$GeV) & $m_{h_2}$ (GeV) & $m_s$ (GeV) & $\Delta a_\mu\times 10^9$ & $\Delta m_W$ (MeV)\\
		\hline
		A & 0.1 & 19 & $3.09$ & $10.5$ & 127 & 98 &2.25  & 70.30\\
		B & 0.6 & 19 & $2.96$ & $10.2$ & 127 & 234 & 2.25 & 70.30 \\
		C & 1.1 & 19 & $2.73$ & $9.69$ & 127 & 316  &2.25  & 70.30 \\
		D & 1.6 & 19 & $2.40$ & $9.13$  & 127 & 380 &2.25  & 70.30 \\
		E & 0.6 & 50 & $2.83$ & $9.55$ & 206 & 236  &2.25  & 70.30 \\
		F & 1.1 & 50 & $2.60$  & $9.08$ & 206 &  317 &2.25  & 70.30 \\
		G & 1.6 & 50 & $2.28$ & $8.64$ & 206 & 382 &2.25  & 70.30 \\
		H & 2.1 & 50 & $1.90$ & $7.92$ & 206 & 437  &2.25  & 70.30\\
		I & 0.6 & 99 & $2.61$ & $8.59$ & 290 & 240 &2.25  & 70.30 \\
		L & 1.1 & 99 & $2.37$ & $8.06$  & 290 & 320 &2.25  & 70.30 \\
		M & 1.6 & 99 & $2.07$ & $7.55$ & 290 & 384 &2.25  & 70.30\\
		N & 2.1 & 99 & $1.73$ & $7.00$ & 290 & 439 &2.25  & 70.30  \\
		O & 2.6 & 99 & $1.40$  & $6.42$ & 290 & 487 &2.25  & 70.30  \\
		\hline
	\end{tabular}
	\caption{Landau pole and perturbativity scales for the  benchmark points in the left plot of  Fig.\,\ref{exclusion plot}. The IR value of the couplings $\lambda_3$ and $\lambda_4$ have been fixed to  $(\lambda_3,\lambda_4)$=$(2\lambda_1,0)$.}
	\label{table:2}
\end{table}

\begin{table}[htbp!]
	\centering
	\begin{tabular}{|p{1.2cm}|p{1.1cm}|p{1.1cm}|p{2.7cm}|p{2.5cm}|p{1.1cm}|p{1.1cm}|p{1.1cm}|p{1.1cm}|}
		\hline
		\multicolumn{9}{|c|}{Benchmark points in Fig.\,\ref{exclusion plot} (Right): $v_\phi=\sqrt{2}\times 200$ GeV} \\
		\hline
		Point  & $\lambda_\phi$ & $\mu_3$ (GeV) & Landau pole\,($\times10^5\,$GeV) & Perturbativity ($\times10^5$\,GeV)& $m_{h_2}$ (GeV) & $m_s$ (GeV)& $\Delta a_\mu\times 10^9$ & $\Delta m_W$ (MeV)\\
		\hline
		A & 0.1 & 19 & $6.32$ & $2.22$ & 146 & 129 & 1.29  & 40.64 \\
		B & 0.6 & 19 & $5.37$ & $1.93 $  & 146 & 311 &1.29 &  40.64\\
		C & 1.1 & 19 & $3.56$ & $1.53$ & 146 & 420 & 1.29 & 40.64  \\
		D & 1.6 & 19 & $1.72$ & $0.870$ & 146 & 507  & 1.29 & 40.64  \\
		E & 0.6 & 50 & $4.81$ & $1.71$ & 238 & 312& 1.29 & 40.64  \\
		F & 1.1 & 50 & $3.16$  & $1.34$ & 238 & 421 & 1.29 & 40.64  \\
		G & 1.6 & 50 & $1.56$ & $0.79$   &  238 & 507 & 1.29 & 40.64 \\
		H & 2.1 & 50 & $0.616$ & $3.09$  & 238 & 581 & 1.29 & 40.64  \\
		I & 0.6 & 100 & $3.87$ & $1.35$ & 336 & 314 & 1.29 & 40.64  \\
		L & 1.1 & 100 & $2.50$ & $1.04$ & 336 & 423  & 1.29 & 40.64  \\
		M & 1.6 & 100 & $1.28$ & $0.65$ & 336 & 509 & 1.29 & 40.64  \\
		N & 2.1 & 100 & $0.541$ & $0.273$  & 336 & 582 & 1.29 & 40.64  \\
		O & 2.6 & 100 & $0.260$  & $0.131$  & 336 & 647 & 1.29 & 40.64 \\
		\hline
	\end{tabular}
  \caption{Landau pole and perturbativity scales for benchmark points in the right plot of  Fig.\,\ref{exclusion plot}. The same choice has been made for $\lambda_3$ and $\lambda_4$. }
\label{table:3}
\end{table}

More generally, in the following subsections we will systematically address issues found that will present themselves in the RG analysis of the benchmark points. This will drive us to a different and somehow restricted region of the parameter space.

\subsection{$Z'$ mass dependence of the Landau pole \label{subsec:landau_pole}}

We advanced in the previous section that the major architect of the low Landau pole is the large IR value of $g_{_{Z'}}$.
The study of the RGEs with $m_{_{Z'}}<500$ GeV, that correspond to lower values of $g_{_{Z'}}$, confirms this expectation. For instance, taking the parameters given in point A for $v_s=200$ GeV and $m_{_{Z'}}=250$ GeV, one finds a singularity at $7.89\times 10^{13}$ GeV: reducing $m_{_{Z'}}$ by a factor $1/2$ pushes the singularity (and with it, the perturbativity scale) 8 order of magnitudes forward. Other representative cases are collected in Table 3, where it is seen that, for $v_s=200$ GeV and $\lambda_\phi=0.1$, the Landau pole can be brought around the Planck scale by taking $m_{_{Z'}}=200$ GeV. For $v_s=150$ GeV and $\lambda_\phi=0.4$, a scenario that is of greater interest in light of the results shown in Fig.\,\ref{anomalies} and of the values of the masses reported in Table \ref{table:2} and Table \ref{table:3}, a Planck scale Landau pole cannot be obtained: further decreases of $m_{_{Z'}}$ do not make any improvement on the results displayed for $m_{_{Z'}}=200$ GeV. We will try in the following to relax some assumptions on the scalar couplings to further increase the pole scale while still fitting the experimental anomalies. We note that besides the above theoretical remarks, values for $m_{_{Z'}}$ smaller than $250\,{\rm GeV}$ are phenomenologically disfavored by the direct $Z'$ searches at the LHC, as discussed in Section \ref{section: pheno_constraints}, and we will not consider them.

Concerning the decrease of the IR value of $m_{_{Z'}}$, it should be noted that the solution to the $W$ boson mass anomaly presented in \cite{Lee:2022nqz}, that, as recalled above, is based on a tree-level correction to the $W$ self-energy due to the $Z$-$Z'$ mixing, requires $m_{_{Z'}}$ to be sufficiently larger than $m_{_Z}$. In this case, in fact, BSM physics brings in a positive correction to the mass, and a region of parameter space where the CDF II measurement can be explained within $1\sigma$ was found \cite{Lee:2022nqz}. On the contrary, in the opposite limit $m_{_{Z'}}\ll m_{_Z}$, the contribution to the $W$ boson mass from BSM physics is negative. Clearly, the findings from CDF II cannot be accommodated with the inverted hierarchy. As can be seen in Table \ref{table:4}$_1$ (see also Fig.\,\ref{anomalies}, where a different value for $\lambda_\phi$ was taken), the correction $\Delta a_\mu$ to the muon $g-2$ and the correction $\Delta m_{_W}$ to the $W$ boson mass are still of the right order to fit the experimental anomalies for $m_{_{Z'}}\gtrsim 250$ GeV, i.e.\,for values of $m_{_{Z'}}$ in agreement with LHC searches for additional $Z$ bosons and below which the Landau pole does not sensibly increase any more. 

\begin{figure}[t]
	\centering
	\includegraphics[scale=0.28]{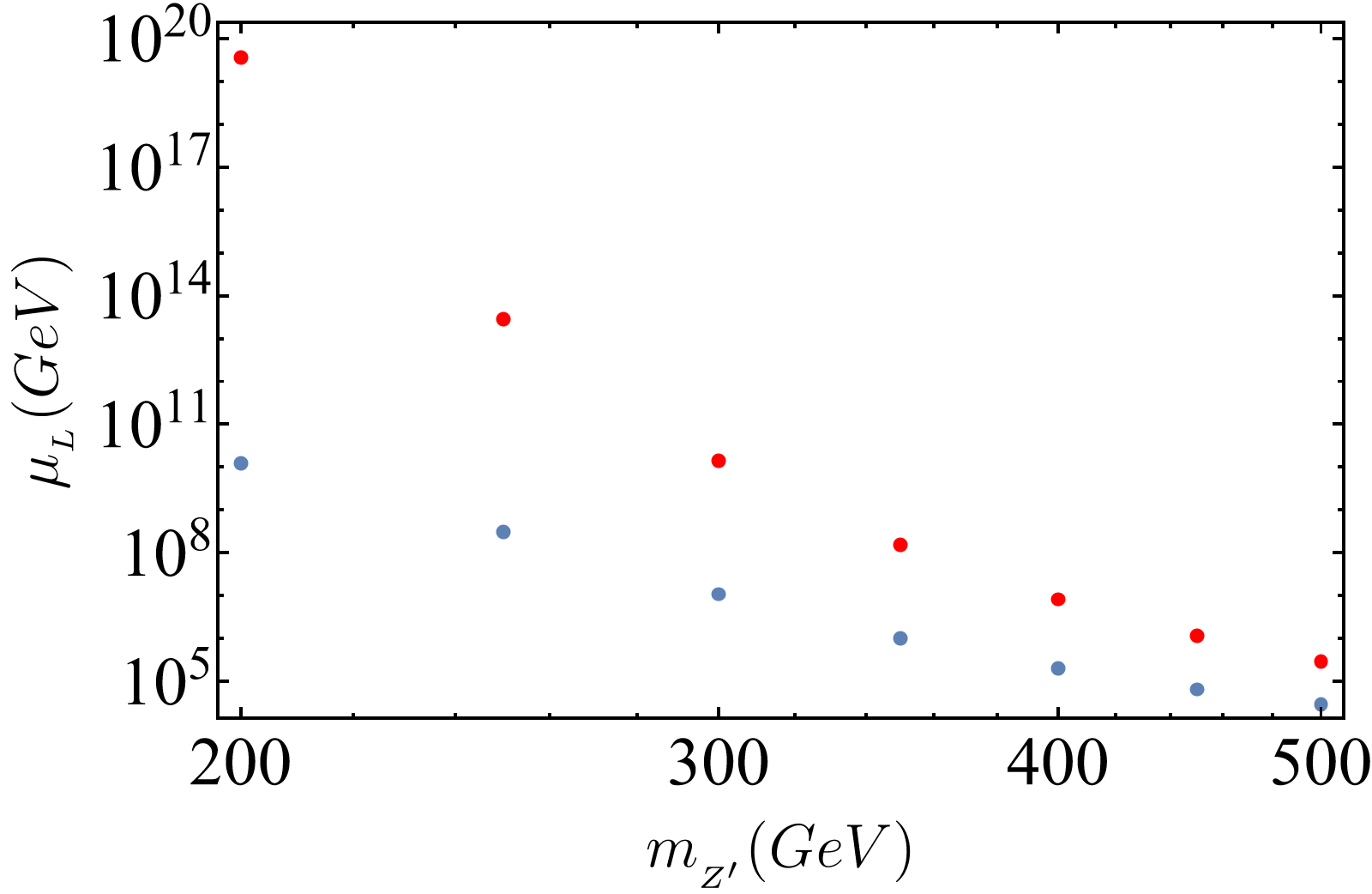} \,\, 
	\includegraphics[scale=0.28]{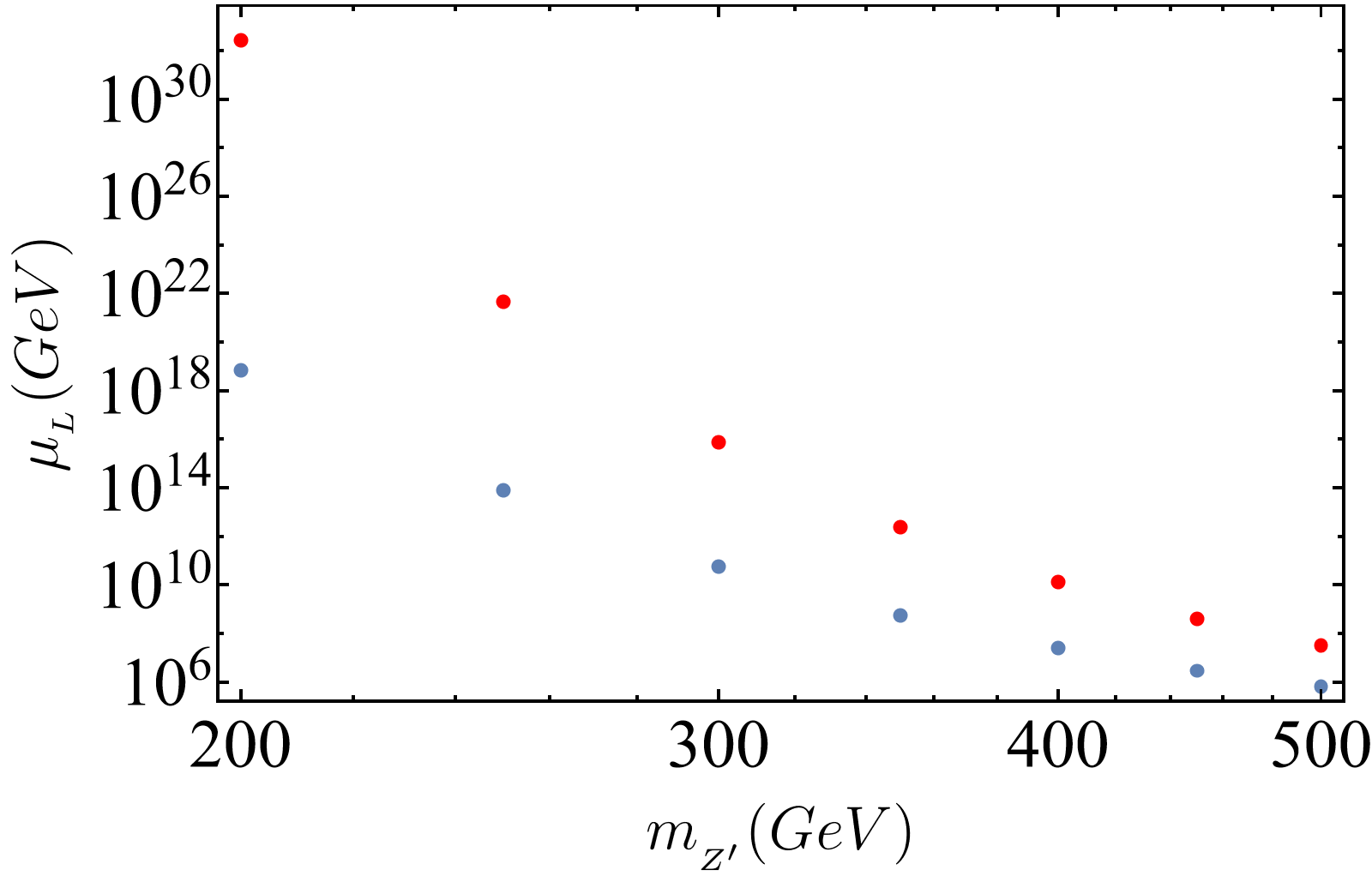}
	\caption{The blue points indicate the Landau pole of the full model, the red ones the Landau pole of Eq.\,\eqref{gZ' RGE} for two sets of parameters in Table \ref{table:4}: the upper table for the left panel and the lower table for the right panel.}
	\label{Landau pole}
\end{figure}

A more quantitative estimate of the role of $g_{_{Z'}}$ in generating the low scale Landau pole can be obtained taking the RGE for $g_{_{Z'}}$, 
\begin{equation}
	\label{gZ' RGE}
\mu \frac{d}{d\mu} g_{_{Z'}}(\mu)=\frac{7}{12\pi^2}g_{_{Z'}}^3(\mu) \quad \longrightarrow \quad  g_{_{Z'}}=\frac{1}{\sqrt{\frac{1}{g^2_{_{Z'}}(\mu_{_{\rm IR}})}-\frac{7}{6\pi^2}\log\left(\frac{\mu}{\mu_{_{\rm IR}}}\right)}},
\end{equation}
and calculating its Landau pole $\mu_{_L}=\mu_{_{\rm IR}}e^{{6\pi^2}/{7g^2_{_{Z'}}(\mu_{_{\rm IR}})}}$ for the different IR values of $g_{_{Z'}}$ in the two sets of the parameters in Table 4. The Landau pole of the full model is compared to $\mu_{_L}$ in Fig.\,\ref{Landau pole}, where it can be seen that the higher the value of $m_{_{Z'}}$ ($g_{_{Z'}}$), the closer the two are. It is easy to conclude from this plot that, for ``high enough" (IR) values of $g_{_{Z'}}$, the Landau pole is almost fully determined by the gauge coupling, that, through its contribution in their beta functions, causes the other couplings to diverge at slightly lower values. Lowering the (IR) value of $g_{_{Z'}}$, the Landau pole of \eqref{gZ' RGE} becomes considerably higher than that of the full model. In this region of parameter space, the appearance of the Landau pole is a more ``collective" phenomenon. All the couplings conspire together to generate it, and the determination of the pole scale from the RG equation of $g_{_{Z'}}$ taken in isolation does not provide a good approximation to the pole scale of the full model any more.

\begin{figure}[t]
	\centering
	\includegraphics[scale=0.55]{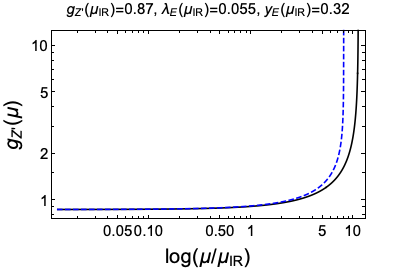} \,\, 
	\includegraphics[scale=0.55]{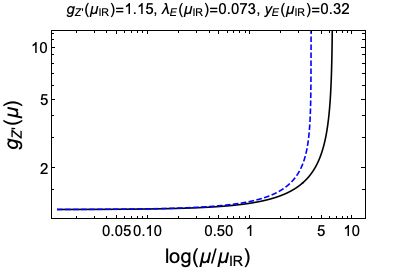}
	\caption{Running $g_{_{Z'}}$ coupling at one-loop and two-loop levels, in black solid and blue dashed lines, respectively. We considered two benchmark models with $(g_{_{Z'}},\lambda_E,y_E)=(0.87,0.055,0.32)$ and $(1.15,0.073,0.32)$, on the left and right plots, respectively. }
	\label{twoloops}
\end{figure}

It is worthwhile to make comments on the effects of two-loop corrections in view of the low scale Landau pole, although a complete two-loop analysis in our model is beyond the scope of our current work. The RG equation for the extra gauge coupling at the two-loop level \cite{twoloops} is given by
\bea
\mu \frac{d}{d\mu} g_{_{Z'}}(\mu)=\frac{7}{12\pi^2}g_{_{Z'}}^3 + \frac{1}{(4\pi)^4} \bigg( 256g_{_{Z'}}^5- 16g_{_{Z'}}^3(\lambda_E^2+2y^2_E)\bigg). 
\eea
Then, ignoring the running of the extra Yukawa couplings, $\lambda_E$ and $y_E$, we get the approximate solution for the running extra gauge coupling at two-loops as
\bea
g_{_{Z'}}(\mu)\simeq\left[ \frac{1}{g^2_{_{Z'}}(\mu_{_{\rm IR}})} -\frac{98\log\left(\frac{\mu}{\mu_{_{\rm IR}}}\right) }{3(28\pi^2+3\lambda^2_E+6y^2_E)}-\frac{96\log\left(\frac{g_{_{Z'}}(\mu) }{g_{_{Z'}}(\mu_{_{\rm IR}})}\right)}{28\pi^2+3\lambda^2_E+6y^2_E} \right]^{-1/2}.
\eea
Then, we get the Landau pole modified to 
\bea
\mu_{_L}\simeq\mu_{_{\rm IR}}\, {\rm exp} \bigg[ \frac{6\pi^2}{7g^2_{_{Z'}}(\mu_{_{\rm IR}})}\Big(1+\frac{3\lambda^2_E}{28\pi^2}+\frac{3y^2_E}{14\pi^2}\Big)-\frac{144}{49}\log\left(\frac{g_{_{Z'}}(\mu_{_L})}{g_{_{Z'}}(\mu_{_{\rm IR}})}\right) \bigg].
\eea
Here, we can take the extra gauge coupling at the Landau pole in the correction term as $g_{_{Z'}}(\mu_{_L})=4\pi$ for which perturbativity breaks down.
For instance, for $\lambda_E=0.055-0.073$ and $y_E=0.32$ in our benchmark models in Section 4.3, we find that the scale of the Landau pole gets reduced sizably at two-loop order due to the two-loop gauge interactions, as compared to the one-loop results.  In the left and right plots in Fig.~\ref{twoloops}, we also depict the running $g_{_{Z'}}$ for the same benchmark models by the numerical analysis at one-loop and two-loop levels, in black solid and blue dashed lines, respectively. As a result, we find that the two-loop corrections are minor far away from the Landau pole, but they become important near the Landau pole. However, the very existence of the Landau pole at a low scale can be identified by the one-loop results, indicating the violation of perturbativity, so we can rely on the one-loop corrections in our RG analysis to extract the meaningful information on the Landau pole. Similar conclusions can be drawn for the perturbativity and stability of the other running couplings such as the Higgs quartic couplings in our model, but we postpone the detailed analysis with two-loop corrections to another work.

\begin{table}[htbp!]
    \centering
    \begin{tabular}{|p{3cm}|p{2cm}|p{2.5cm}|p{2.45cm}|p{1.8cm}|p{2.1cm}|}
			\hline
			\multicolumn{6}{|c|}{Parameter set: $v_\phi=\sqrt{2}\times 150$ GeV, $\lambda_{\phi}=0.4$, $\mu_3=50$ GeV, $\sin\beta=0.18$} \\
			\hline
			$m_{_{Z'}}$(GeV) $\rightarrow g_{_{Z'}}$ &  LP (GeV) & Pert (GeV) & VSB (GeV) &  $\Delta a_\mu\times 10^9$ & $\Delta m_W({\rm MeV})$ \\  
			\hline
			500 $\rightarrow$ 1.15 & $2.90\times10^4$  & $9.71\times10^3$ &  $-$  & 2.25& 70.30\\
			450 $\rightarrow 1.04$ & $6.51\times 10^4$ & $2.18\times 10^4$ &  $-$ & 2.15 & 70.25   \\
			400 $\rightarrow 0.92$ & $2.02\times10^5$ & $6.86\times10^4$ &  $-$ & 2.06 & 70.17 \\
			350 $\rightarrow 0.81$ & $1.02 \times 10^6$ &  $3.59 \times 10^5$ &   $-$ & 1.97 & 70.01  \\
			300 $\rightarrow 0.69$ & $1.09\times 10^{7}$ & $4.07\times10^{6}$ &  $-$ &  1.88 & 69.66  \\
			250 $\rightarrow 0.58$ & $3.05\times 10^{8}$& $1.36\times 10^{8}$ & $-$ & 1.80 & 68.75 \\
			200 $\rightarrow 0.46$ & $1.22\times 10^{10}$ & $7.83\times 10^{9}$ & $2.50\times 10^8$ & 1.72 &  68.61\\
			\hline
		\end{tabular}\\ \vspace{0.2cm} 
	\begin{tabular}{|p{3cm}|p{2cm}|p{2.5cm}|p{2.45cm}|p{1.8cm}|p{2.1cm}|}
		\hline
		\multicolumn{6}{|c|}{Parameter set: $v_\phi=\sqrt{2}\times 200$ GeV, $\lambda_{\phi}=0.1$, $\mu_3=19$ GeV, $\sin\beta=0.18$} \\
		\hline
		$m_{_{Z'}}$(GeV) $\rightarrow g_{_{Z'}}$ &  LP (GeV) & Pert (GeV) & VSB (GeV) & $\Delta a_\mu\times 10^9$ & $\Delta m_W({\rm MeV})$ \\  
		\hline
		500 $\rightarrow$ 0.87 & $6.32\times10^5$  & $2.22\times10^5$ & $-$ & $1.29$ &  40.64 \\
		450 $\rightarrow 0.79$ & $2.91\times 10^6$ & $1.04\times 10^6$  & $-$ & $1.23$  & 40.61 \\
		400 $\rightarrow 0.70$ & $2.46\times10^7$ & $9.15\times10^6$  & $-$ & $1.18$ & 40.56 \\
		350 $\rightarrow 0.61$ & $5.32 \times 10^8$ &  $2.08 \times 10^8$ & $1.35\times 10^8$ & $1.13$  & 40.47 \\
		300 $\rightarrow 0.52$ & $5.30\times 10^{10}$ & $2.23\times10^{10}$ & $1.19 \times 10^8$  & $1.08$ & 40.27 \\
		250 $\rightarrow 0.44$ & $6.43\times 10^{13}$& $3.10\times 10^{13}$ & $1.59 \times 10^8$  & $1.03$  &  39.75\\
		200 $\rightarrow 0.35$ & $3.08\times 10^{18}$ & $1.82\times 10^{18}$ & $2.49\times 10^8$ & $0.99$  & 37.94 \\
		\hline
	\end{tabular}
  \caption{Landau pole (LP), Perturbativity (Pert) and Vacuum stability (VSB) scales for some $Z'$ masses with two examples for the parameter set. In the last two columns, we also showed the numerical values of the corrections to the muon $g-2$ and the $W$ boson mass in our model. }
    \label{table:4}
\end{table}

\begin{figure}[t]
	\centering
\includegraphics[scale=0.4]{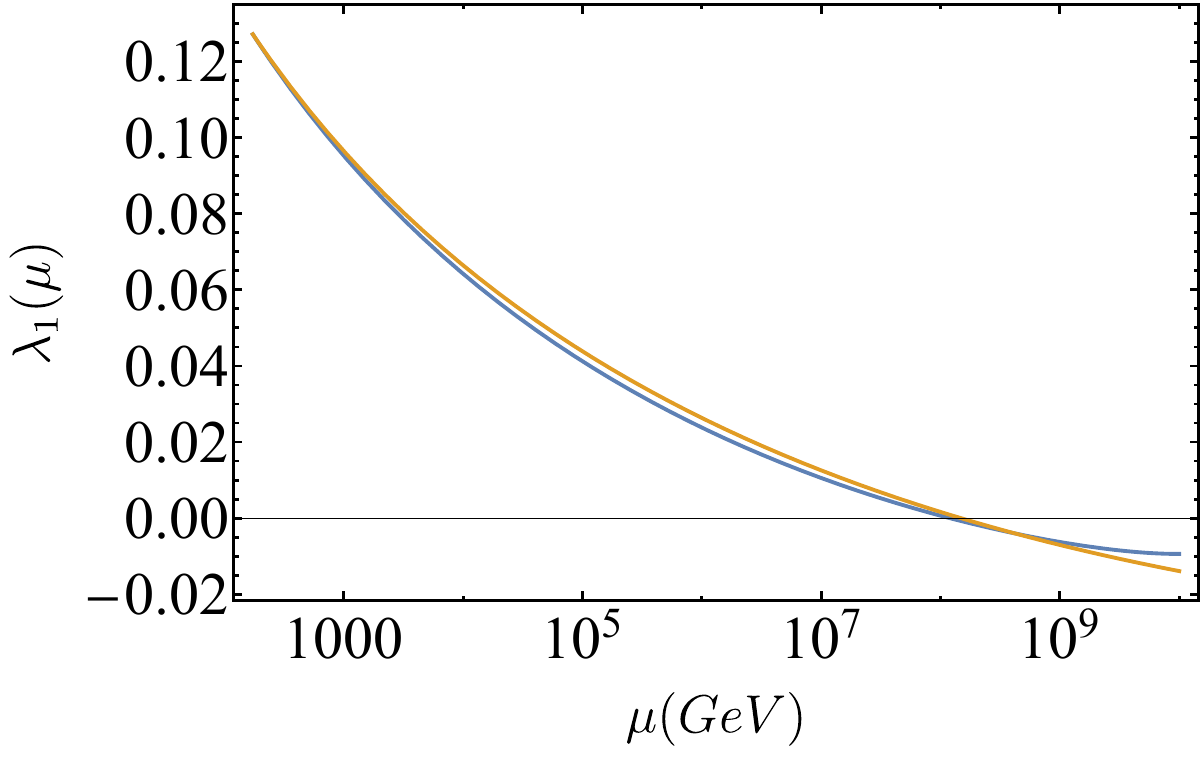}
\includegraphics[scale=0.4]{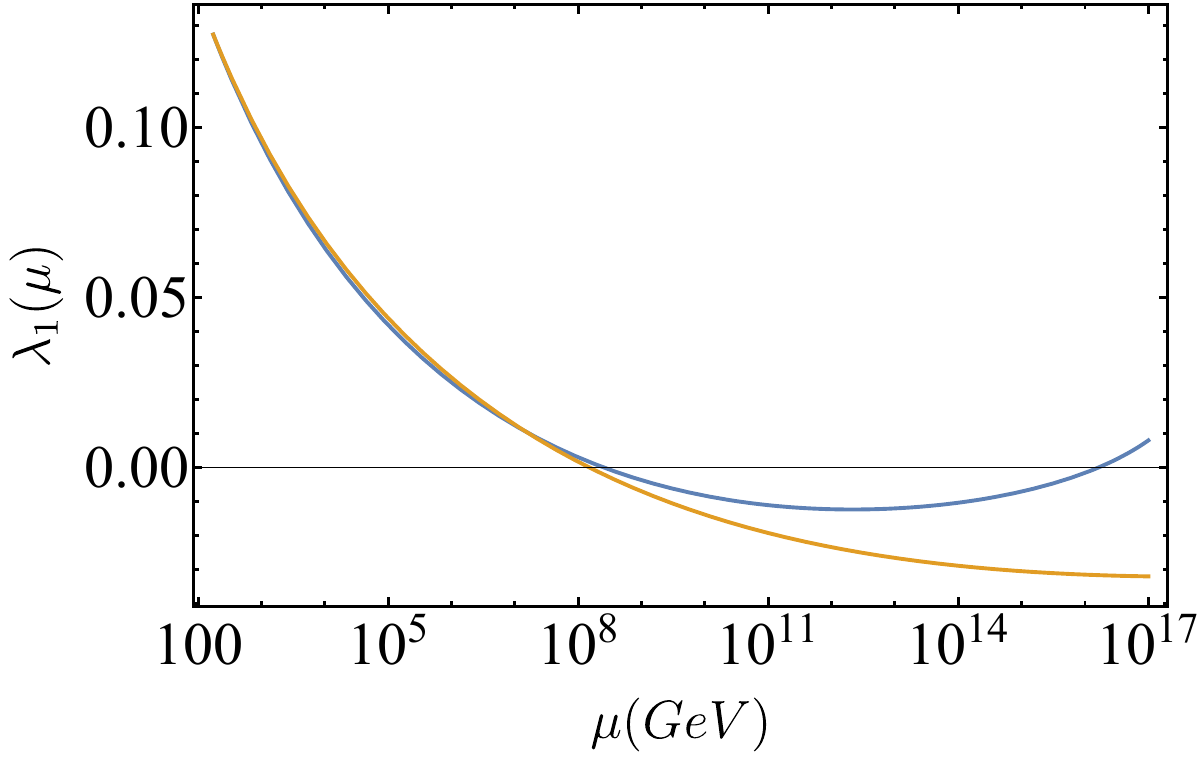}
	\caption{The blue curve indicates the flow of $\lambda_1$ for the choice of parameters corresponding to point A in Table 3 and with $m_{_{Z'}}=300$ GeV (\textit{left panel}) $m_{_{Z'}}=200$ GeV (\textit{right panel}). The yellow curve corresponds to the flow of $\lambda_1$ in the Standard Model.}
	\label{fig lambda1}
\end{figure}

When the pole scale is increased above $\sim 10^8$ GeV, the range of the flow is large enough to see the appearance of an instability of the scalar potential. As in the SM, this is due to the fact that the Higgs-like quartic coupling $\lambda_1(\mu)$ becomes negative while running toward high energies\footnote {The other stability conditions \eqref{conditions stability}, \eqref{condition stability 2} are respected with the running couplings for all (admissible) values of $\mu$.}. The scale where such an instability appears is indicated for some representative cases in Table 4. This should be compared to the corresponding case of the SM, where, with our parametrization, $\lambda(\mu)=0$ for $\mu\sim1.48\times 10^8$ GeV at the one-loop level\footnote{With our parametrization, the SM quartic coupling vanishes for $\mu \sim 1.25\times 10^{10}$ GeV at the two-loops.}. The shift in the IR boundary value of the top coupling caused by the $\cos\beta$ factor in $m_t=y_t v \cos\beta/\sqrt2$ competes with the presence of several additional bosonic states in such a way that, with the IR boundary values chosen in Table \ref{table:4}, the instability scale does not change much with respect to the SM one.

In Fig.\,\ref{fig lambda1},  a comparison between the flow of $\lambda_1(\mu)$ in our model and the flow of $\lambda(\mu)$ in the SM can be found for $v_s= 200$ GeV, $\lambda_\phi=0.1$,  $\mu_3=19$ GeV and $m_{_{Z'}}=300$ GeV (left panel), $m_{_{Z'}}=200$ GeV (right panel). This choice of parameters allow to have a larger domain for the the RG evolution. The behaviour of $\lambda_1(\mu)$ just discussed goes against the naive expectation that the presence of (several) additional bosonic states automatically improves stability; there is rather a delicate competition between the enhanced top quark contribution and the BSM ones.  As we will see below, radiative stability of the potential can be realized, but the positive shift in the top coupling caused by the enlargement of the scalar sector makes it a not too trivial matter.       
It should also be noted that, thanks to the additional bosonic degrees of freedom, after crossing zero $\lambda_1(\mu)$ soon stops its descent and starts to grow faster than in the SM. In the case depicted in Fig.\,\ref{fig lambda1}$_2$, for instance, the coupling becomes positive again for $\mu\gtrsim 10^{16}$ GeV before reaching its Landau pole at $\mu\sim 3\times 10^{18}$ GeV.

\section{General alignment limits in the IR}

We extend the RG analysis to the cases where the doublets are not completely decoupled from the scalar singlet, and their quartic couplings do not necessarily realize alignment.

\subsection{Detuning quartic couplings}

So far, we followed \cite{Lee:2022nqz} in the determination of the conditions for alignment. Namely, we took the condition $\lambda_1=\lambda_2=\lambda_{34}/2$ for the IR parameters that, as stressed above, was shown in \cite{Pilaftsis:2016erj} to correspond to the maximally symmetric configuration for the scale-invariant sector of the classical potential $V(\phi, H, H')$ in \eqref{scalarpot}. 
As already observed in \cite{Lee:2022nqz}, however, this is only a particular solution that realizes the alignment of the two Higgs doublets. Below, we derive the general solution for alignment imposing as a constraint that the lowest eigenvalue of the $2\times2$ mass matrix $\widetilde{\mathcal M}^2_H$ \eqref{Mtilda} coincides with the physical Higgs boson mass. 

Taking $\widetilde{\mathcal M}^2_H$, that we rewrite here for the ease of the reader,
\begin{eqnarray*}
	\widetilde {\mathcal M}^2_H =\left(
	\begin{array}{cc} 2\lambda_1 v_1^2 +\frac{\mu_3}{\sqrt 2}\tan\beta \,v_\phi &  \lambda_{34} v_1 v_2-\frac{\mu_3}{\sqrt 2}v_\phi \vspace{0.2cm} \\ \lambda_{34} v_1 v_2-\frac{\mu_3}{\sqrt 2}v_\phi & 2\lambda_2 v_2^2 +\frac{\mu_3}{\sqrt 2}\cot\beta \,v_\phi  \vspace{0.2cm} 
	\end{array} \right), 
\end{eqnarray*}
and applying a rotation of angle $\beta$ parametrized by the matrix $S_\beta$, we obtain 
\begin{eqnarray}
	\label{M tilde beta}
\widetilde {\mathcal M}^2_{H,\,\beta}\equiv S_\beta^{-1}\widetilde{\mathcal M}^2_H S_\beta =\left(
	\begin{array}{cc} A  &  C \vspace{0.2cm} \\ C & B  \vspace{0.2cm} 
	\end{array} \right), 
\end{eqnarray}
with 
\begin{align}
A&=2\left(\lambda_1\cos^4\beta+\lambda_{34}\cos^2\beta \sin^2\beta+\lambda_2\sin^4\beta\right)v^2, \\
B&=\frac{1}{\sqrt 2}\frac{\mu_3 v_\phi}{\sin\beta \cos\beta}+2\sin^2\beta \cos^2\beta \left(\lambda_1+\lambda_2-\lambda_{34}\right)v^2, \\
C&= \left(\left(\lambda_{34}-2\lambda_1\right) \cos^3\beta \sin\beta+\left(2\lambda_2-\lambda_{34}\right)\sin^3\beta\cos\beta\right) v^2.
\end{align}
In the following, we denote with $h$ and $h'$ the two mass eigenstates obtained after diagonalization:
\begin{eqnarray}
	\left(
	\begin{array}{c} h  \vspace{0.2cm} \\ h' \vspace{0.2cm}
	\end{array}\right) = S^{-1}_\beta 	\left(\begin{array}{c} h_1  \vspace{0.2cm} \\ h_2 \vspace{0.2cm}
	\end{array} \right).
\end{eqnarray}
In terms of $\lambda_{34}$, the general solution to the equation $C=0$ reads
\begin{equation}
	\label{sol l3}
\lambda_{34}= 2\frac{\lambda_1-\lambda_2\tan^2\beta}{1-\tan^2\beta}.
\end{equation}  
For later convenience, it is useful to also solve \eqref{sol l3} in terms of $\tan \beta$ as (this is only well defined when $2\lambda_2-\lambda_{34}\ne 0$)
\begin{equation}
\tan^2\beta =\frac{2\lambda_1-\lambda_{34}}{2\lambda_2-\lambda_{34}}. 
\end{equation}
Moreover, for $A$ to be the SM Higgs squared mass $m_h^2=2\lambda_{\rm SM} v^2$, we need\footnote{For consistency, this should be accompanied by 
	\begin{equation}
	\frac{\mu_3v_\phi}{\sqrt 2\sin\beta\cos\beta}\ge \left(\lambda_1+\lambda_2+\frac{\lambda_1-\lambda_2}{\cos 2\beta}\right),
\end{equation}
otherwise the second Higgs boson $h'$ is lighter than the SM Higgs.} $\lambda_1\cos^4\beta+\lambda_3\cos^2\beta \sin^2\beta+\lambda_2\sin^4\beta=\lambda_{\rm SM}$, where $\lambda_{\rm SM}$ is the Standard Model Higgs quartic coupling. In passing, we note that this condition also guarantees that the three and four point vertices for the Higgs boson $h$ obtained after diagonalization are the same as in the SM. Solving for $\lambda_2$, we get
\begin{equation}\label{constraint l2}
\lambda_2=\lambda_1+\left(\lambda_1-\lambda_{\rm SM}\right)\frac{\cos2\beta}{\sin^4\beta}.
\end{equation}
Finally, inserting \eqref{constraint l2} in \eqref{sol l3}, the latter becomes 
\begin{equation}\label{constraint l3}
\lambda_{34}=2\lambda_1-2\,\frac{\lambda_1-\lambda_{\rm SM}}{\sin^2\beta}.
\end{equation}

These last equations are quite interesting, as they tell us how much a detuning of $\lambda_1$ from $\lambda_{\rm SM}$ affects $\lambda_2$ and $\lambda_{34}$ if we keep the SM Higgs in the alignment limit. 
It is immediate to realize that, for small $\beta$ (as required by the $W$ boson mass as well as LHC bounds), the coefficient $\cos2\beta/\sin^4\beta$ in front of $\lambda_1-\lambda_{\rm SM}$ in \eqref{constraint l2} (the coefficient $1/\sin^2\beta$ as well, although it is obviously smaller) is very large. For instance, for $\sin\beta =0.18$ we have  $\cos2\beta/\sin^4\beta\sim 891$ and $1/\sin^2\beta\sim 31$. Needless to say, ``large" values of the scalar couplings are unacceptable, as they easily result in too low Landau pole scales. For instance, a $0.47\%$ relative detuning of $\lambda_1$ with respect to the last example shown in Table \ref{table:4}$_2$, $\lambda_1-\lambda_{\rm SM}=0.0006$ (we take $\lambda_{\rm SM}=0.1272$ at the top scale $m_t$), results in a $421\%$ relative detuning of $\lambda_2$, $\lambda_2-\lambda_{\rm SM}=0.54$, a $14\%$ relative detuning of $\lambda_{34}$, $\lambda_{34}-2\lambda_{\rm SM}=-0.04$, and  brings the Landau pole down from $\mu_L\sim 3\times 10^{18}$ GeV to $\mu_L\sim 7\times 10^{7}$ GeV. 

The extreme sensitivity of the couplings, especially of $\lambda_2$, to the detuning of $\lambda_1$ from $\lambda_{\rm SM}$ tells us that the constraint brought by eq.\,\eqref{constraint l2} is extremely tight. Detuning $\lambda_1$ to larger values $\lambda_1>\lambda_{\rm SM}$ might stabilize the vacuum, but we see here that it does not come without cost. For benchmark points as the last one in Table \ref{table:4}$_1$, one can device a small enough detuning to shift the $\lambda_1(\mu)$ curve all in the upper half plane up to a Landau pole that is sufficiently far from the SM (one-loop) instability scale $\mu\sim 10^8$ GeV, making the indication of one-loop vacuum stability quite trustable. However, the same cannot be done for benchmarks of phenomenological interest, such as the $m_{_{Z'}}\gtrsim250$ GeV ones in Table \ref{table:4}$_2$. There is in fact too little a hierarchy between the Landau pole in the tuned scenario and the instability scale (when it exists) to shift the couplings and obtain truthful indications for stability.    

Before closing this section, we should note that \eqref{constraint l2}, as well as a slight generalization of \eqref{constraint l3}, is not restricted to the present model, but is of more general validity for 2-Higgs doublet models. In $\widetilde{\mathcal M}^2_S$, in fact, the dark Higgs field $\phi$ only accounts for the presence of $\mu_3$, from which $A$ and $C$ are independent. More precisely, after a $\mathbb Z_2$ symmetry $H_1\to H_1,\, H_2\to - H_2$ is imposed and a real small breaking parameter $\mu_{12}$ is inserted, the inert 2-Higgs doublet model's potential reads
\begin{align}
V&=\mu_1^2|H_1|^2+\mu_2|H_2|^2-\mu_{12}^2\left(H^\dagger_1 H_2+\,{\rm h.c.}\right)+\lambda_1|H_1|^4+\lambda_2|H_2|^4+\lambda_3|H_1|^2|H_2|^2\nonumber \\
&+\lambda_4 |H_1^\dagger H_2|^2+\frac{\lambda_5}{2}\left((H_1^\dagger H_2)^2+(H_2^\dagger H_1)^2\right).
\end{align}
Using the minimization conditions, that for a normal vacuum 
\begin{equation}
H_1=\frac{1}{\sqrt 2}	\left(
\begin{array}{c} 0 \vspace{0.2cm} \\ v_1 \vspace{0.2cm}
\end{array}\right), \qquad H_2=\frac{1}{\sqrt 2}	\left(
\begin{array}{c} 0 \vspace{0.2cm} \\ v_2 \vspace{0.2cm}
\end{array}\right) 
\end{equation}
with $v_2/v_1=\tan\beta$, 
amount to 
\begin{equation}
\begin{cases}
\mu_1^2+\lambda_1 v_1^2+\frac{\lambda_3+\lambda_4+\lambda_5}{2}v_2^2= \mu_{12}^2\frac{v_2}{v_1},\\
\mu_2^2+\lambda_2 v_2^2+\frac{\lambda_3+\lambda_4+\lambda_5}{2}v_1^2= \mu_{12}^2\frac{v_1}{v_2},
\end{cases}
\end{equation}
and defining $\lambda_{345}\equiv \lambda_3+\lambda_4+\lambda_5$, the neutral CP-even mass matrix reads, in the vacuum,
\begin{eqnarray*}
	 \mathcal M^2 =\left(
	\begin{array}{cc} 2\lambda_1 v_1^2 +\mu_{12}^2\tan\beta  &  \lambda_{345} v^2\sin\beta\cos\beta-\mu_{12}^2 \vspace{0.2cm} \\ \lambda_{345} v^2\sin\beta\cos\beta-\mu_{12}^2 & 2\lambda_2 v_2^2 +\mu_{12}^2\cot\beta  \vspace{0.2cm} 
	\end{array} \right). 
\end{eqnarray*} 
The above matrix is the same as $\widetilde{\mathcal M}^2_S$ in \eqref{Mtilda} with $\mu_3 v_\phi/\sqrt 2$ replaced by $\mu_{12}^2$ and $\lambda_{34}$ replaced by $\lambda_{345}$. A rotation by an angle $\beta$ would then produce the same matrix as \eqref{M tilde beta}, with replacements in $A$, $B$ and $C$ similar to those discussed above. The equations $A=2\lambda_{\rm SM} v^2$ and $C=0$ then result in the conditions
\begin{equation}
	\label{sol 2HDM}
\begin{cases}
\lambda_{345}=2\lambda_1-2\,\frac{\lambda_1-\lambda_{\rm SM}}{\sin^2\beta}\\
\lambda_2=\lambda_1+\left(\lambda_1-\lambda_{\rm SM}\right)\frac{\cos2\beta}{\sin^4\beta},
\end{cases}
\end{equation} 
that provide only a slight generalization to \eqref{constraint l2} and \eqref{constraint l3}. The consequences, of course, are the same.

\subsection{General decoupling of the singlet scalar}
\label{subsec: general_decoupling}

Acceptable values of the additional bosons' physical masses are obtained only with sufficiently large values of $\mu_3$. As $\cot\beta\sim 5.5$, large values of $\mu_3$ can easily results in dangerously large IR values for $\lambda_{H'\phi}$. 

The decoupling scenario of \cite{Lee:2022nqz}, where the scalar field $s$ is decoupled from $h_1$ and $h_2$, can actually be relaxed to a more general scenario, within which large values of $\mu_3$ can be more easily accommodated, and the Landau poles found for $v_s=150$ GeV can be slightly uplifted. Taking the $3\times 3$ scalar mass matrix $\mathcal M^2_H$ \eqref{scalar mass matrix} in the vacuum, in fact, we can first diagonalize its $2\times 2$ upper part ($\widetilde{\mathcal M}^2_H$), which results in the change of basis $(h_1,h_2)\to (h,h')$, and then require the scalar $s$ to be decoupled from the physical Higgs boson $h$. Borrowing the notation of Appendix A for the components of $\mathcal M^2_H$, after a rotation of angle $\alpha$ in the $(h_1,h_2)$ subspace the mass matrix $\mathcal M^2_H$ reads
\begin{eqnarray}
	\mathcal M^2_H=\left(
	\begin{array}{ccc}
		A & B & C \\
		B & D & E \\
		C & E & F \\
	\end{array}
	\right),
\end{eqnarray}
with
\begin{eqnarray}
\label{A}	
A &=& \frac{1}{2} \left(M^2_{H,\,11}+M^2_{H,\,22}+(M^2_{H,\,11}-M^2_{H,\,22}) \cos 2 \alpha +2 M^2_{H,\,12} \sin2\alpha \right), \\
\label{B}
B &=& \frac12\left(M^2_{H,\,22}-M^2_{H,\,11}\right) \sin2\alpha+M^2_{H,\,12} \cos 2\alpha,  \\
C &=&  M^2_{H,\,13}\cos\alpha+M^2_{H,\,23} \sin \alpha,\\ 
D &=& \frac{1}{2} \left(M^2_{H,\,11}+M^2_{H,\,22}-(M^2_{H,\,11}-M^2_{H,\,22}) \cos 2 \alpha -2 M^2_{H,\,12} \sin2\alpha \right),  \\
\label{E}
E &=&  M^2_{H,\,23} \cos\alpha-M^2_{H,\,13} \sin\alpha,\\
F &=&  M^2_{H,\,33}. 
\end{eqnarray}
As it is well-known, the diagonalization condition $B=0$ reads
\begin{equation}
\tan2\alpha=-2 \frac{M^2_{H,\,12}}{M^2_{H,\,22}-M^2_{H,\,11}},
\end{equation}
and $A$ must be $A=m_h^2$, with $m_h$ the Higgs mass.  
Decoupling the Higgs $h$ from $s$ amounts to require
\begin{equation}
	\label{semi dec general}
M^2_{H,\, 13}\cos\alpha+M^2_{H,\, 23}\sin\alpha=0. 
\end{equation}
Plugging the explicit expressions of $M^2_{H,\, 13}$ \eqref{M2S 13} and $M^2_{H,\, 23}$ \eqref{M2S 23} in \eqref{semi dec general}, we obtain
\begin{equation}
	\label{semidec couplings}
\left(\lambda_{H\phi} \cos\alpha \cos\beta+\lambda_{H'\phi}\sin\alpha\sin\beta\right) v_\phi =\frac{\mu_3}{\sqrt 2} \left(\cos\alpha\sin\beta+\sin\alpha\cos\beta\right). 
\end{equation} 
In the alignment limit $\alpha=\beta$, the above expression simplifies to 
\begin{equation}
	\label{semi dec aligned}
\left(\lambda_{H\phi} \cos^2\beta+\lambda_{H'\phi}\sin^2\beta\right) v_\phi =\sqrt 2 \,\mu_3 \sin\beta\cos\beta.
\end{equation}
It is immediate to verify that $\lambda_{H\phi}=\mu_3\tan\beta/(\sqrt 2\, v_\phi)$ and $\lambda_{H'\phi}=\mu_3\cot\beta/(\sqrt 2\, v_\phi)$ give a specific solution of \eqref{semi dec aligned} that also realize $E=0$, as it should. 

With the conditions above, the ``semi-diagonalized" matrix becomes 
\begin{equation}
\left(
\begin{array}{ccc}
\lambda^- & 0 & 0 \\
0 & \lambda^+ &  \frac{M^2_{H,\,23}}{\cos\alpha} \\
0 & \frac{M^2_{H,\,23}}{\cos\alpha} & M^2_{H,\,33} \\
\end{array}
\right),
\end{equation}
with 
\begin{equation}
\lambda^\pm= \frac{1}{2} \left(M^2_{H,\,11}+M^2_{H,\,22}\pm\left(M^2_{H,\,22}-M^2_{H,\,11}\right) \sqrt{\frac{(M^2_{H,\,22}-M^2_{H,\,11})^2+4 M^4_{H,\,12}}{(M^2_{H,\,22}-M^2_{H,\,11})^2}}\right).
\end{equation}
As we want the Higgs to be the lightest scalar, we also require $\lambda^-<\lambda^+$. This is satisfied when $M^2_{H,\,22}>M^2_{H,\,11}$, in which case the above expressions for $\lambda^\pm$ easily simplify. Finally, the eigenvalues of the $2\times 2$ lower mass matrix will give the physical masses of the additional scalars as 
\begin{equation}
	\label{mplusminus}
m^2_\pm=\frac{1}{2} \left(\lambda^+ +M^2_{H,\, 33}\pm\sqrt{\left(\lambda^+ -M^2_{H,\, 33}\right)^2+4 \frac{M^4_{H,\, 23}}{\cos^2\alpha}}\right).
\end{equation}

\begin{figure}
\centering
\includegraphics[scale=0.5]{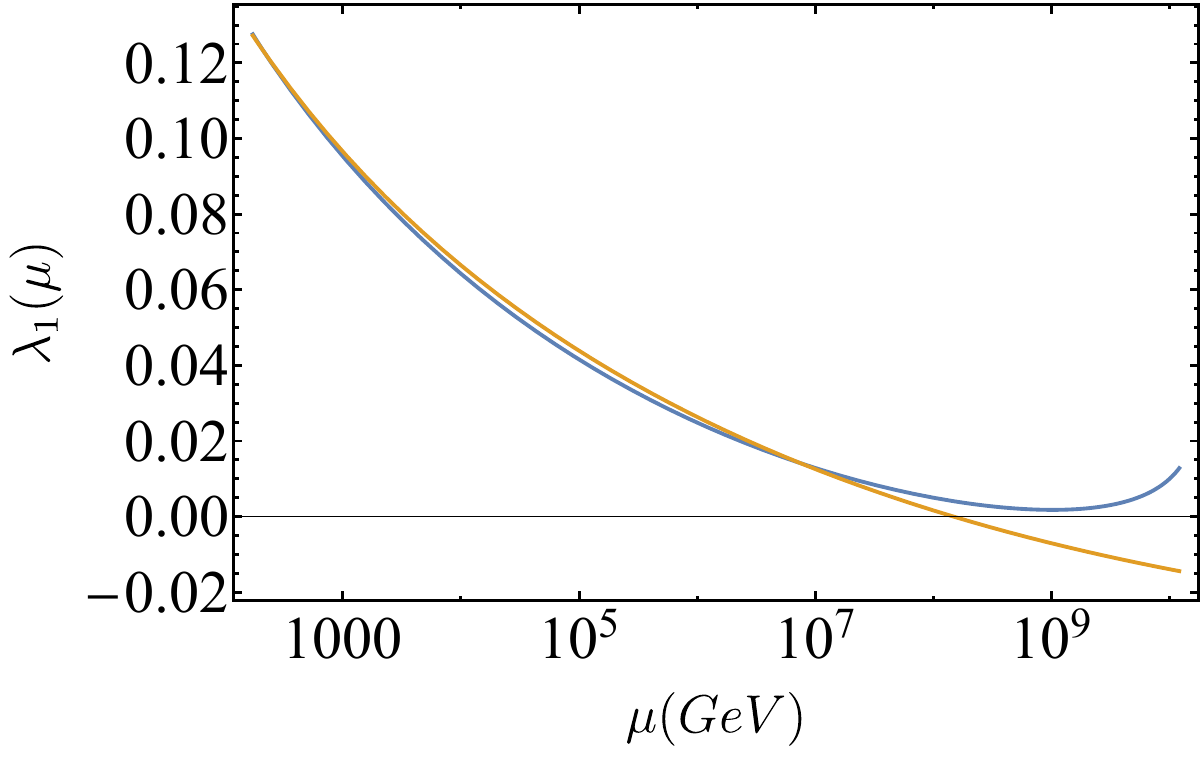}
\caption{Comparison between the the flow of $\lambda_1(\mu)$ (blue curve) and the flow of $\lambda(\mu)$ in the Standard Model (yellow curve) for the benchmark point with $m_{_{Z'}}=250$ GeV in Table \ref{table:4}$_1$ in the detuned scenario with IR boundary values $\lambda_1=0.12746$, $\lambda_4=0$, $\lambda_{H'\phi}=-2\sqrt{\lambda_2\lambda_\phi}\sim-0.758$, $\lambda_{H\phi}=\sqrt 2\frac{\mu_3\tan\beta}{v_\phi}-\lambda_{H'\phi}\tan^2\beta\sim0.086$ and $\lambda_2,\lambda_3$ determined from \eqref{constraint l2} and \eqref{constraint l3}, respectively.}
\label{lambda1detuned}
\end{figure}

Using the general solution for decoupling described in this section, we can now study the RG flow of the parameters from the benchmark point with $m_{_{Z'}}=250$ GeV in Table \ref{table:4}$_1$ as we take different solutions of eq.\,\eqref{semi dec aligned} for $\lambda_{H\phi}$ and $\lambda_{H'\phi}$. The numerical investigation indicate that higher values for the Landau pole are obtained for higher values of $\lambda_{H\phi}$. In particular, the highest possible value is found combining eq.\,\eqref{semi dec aligned} with the stability conditions \eqref{conditions stability}, \eqref{condition stability 2}, in particular saturating the inequality $\lambda_{H'\phi}+2\sqrt{\lambda_2\lambda_\phi}\ge 0$. For the $(\lambda_{H\phi},\lambda_{H'\phi})$ couple determined in this way, we find $\mu_L\sim2.18\times 10^9$ GeV and an instability scale $\mu_{i}\sim 1.03\times 10^8$ GeV for $\lambda_1=\lambda_{\rm SM}$. A detuning of $\lambda_1$ then allows to find points that generate Landau pole scales $\mu_L\lesssim 10^9$ GeV and a one-loop stable potential up to $\mu_L$. For instance, we find $\mu_L\sim 9.69\times 10^8$ GeV, with a perturbativity scale $\mu_P\sim 5.47\times 10^8$ GeV, and the flow reported in Fig.\,\ref{lambda1detuned} for a $0.2\%$ detuning of $\lambda_1$ ($\lambda_1=0.12746$). Another convenient choice for $\lambda_{H\phi}$ and $\lambda_{H'\phi}$ might be $\lambda_{H\phi}=\lambda_{H'\phi}$, for which we find $\mu_L\sim 1.33\times 10^9$ GeV in the tuned scenario ($\lambda_1=\lambda_{\rm SM}$). The differences between the two cases are too small to be considered as significant ones within our one-loop investigation, whose aim is rather to assess the viability of our model with respect to its flow towards the UV, to find the approximate scale at which new physics should complete it, and understand whether it can lead to a (more) stable potential or not. We will see, in the next section, that the second choice is actually more interesting as it will leave us more freedom to explore the parameter space of the theory when the two doublets are not aligned.   

\subsection{Breaking the alignment}

We comment in this section on the possibility of relaxing the assumption that the two doublets are in the aligned limit. We consider, instead, the $(h_1,h_2)$ mass matrix $\widetilde{\mathcal M}^2_H$ to be diagonalized by a generic angle in the vacuum. 

To this end, we perform, as in Section \ref{subsec: general_decoupling}, a rotation of angle $\alpha$ in the $(h_1,h_2)$ subspace, and impose the following conditions: (i) the resulting matrix is diagonal in its $2\times2$ upper left component, that is $B=0$ in \eqref{B}; (ii) the resulting Higgs field $h$ is decoupled from the dark Higgs, that is $C=0$ in \eqref{E}; and (iii) the $(1,1)$ component of the matrix obtained after diagonalization is equal to $m_h^2$, that is $A=m_h^2$ in \eqref{A}.
The second condition simply gives eq.\,\eqref{semidec couplings}. The other two conditions can be solved for $\lambda_2$ and $\lambda_{34}$, for instance, and give 
\begin{eqnarray}
\label{misaligned l2}
\lambda_2&=& \lambda_1 \frac{\sin^2\alpha-\cos^2\alpha\sin^2\beta}{\sin^2\alpha \sin^2\beta}+ \left(\lambda_1-\lambda_{\rm SM}\right)\frac{\cos2\alpha}{\sin ^2\alpha \sin^2\beta}+\frac{\mu_3 v_\phi}{4 v^2}\frac{\cos2\alpha-\cos2\beta}{\sin^2\alpha \sin^3\beta\cos\beta},\\
\label{misaligned l34}
\lambda_{34}&=&-\frac{2v^2 \cot\alpha (\lambda_1-\lambda_{\rm SM}-\lambda_1\sin^2\beta)+\mu_3 v_\phi (\cot\alpha\tan \beta -1)}{v^2 \sin\beta \cos\beta}.
\end{eqnarray}
It is immediate to verify that the two equations above reduce to \eqref{constraint l2} and \eqref{constraint l3}, respectively, when $\alpha=\beta$. Equations \eqref{misaligned l2} and \eqref{misaligned l34} show an interesting interplay between  the quartic coupling $\lambda_1$ and the diagonalization angle $\alpha$. As we will see, a misalignment in the doublets can be used to tame the enhancement found in \eqref{constraint l2} and \eqref{constraint l3} for detuned quartic couplings $\lambda_1\ne \lambda_{\rm SM}$. 

\begin{figure}
\centering
\includegraphics[scale=0.5]{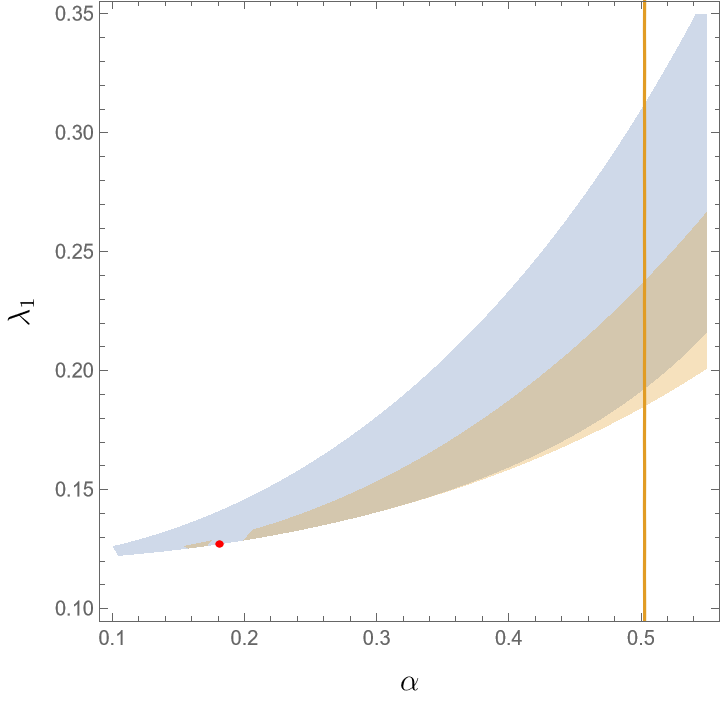}
\,\,\,\,\,\,\,\,
\includegraphics[scale=0.5]{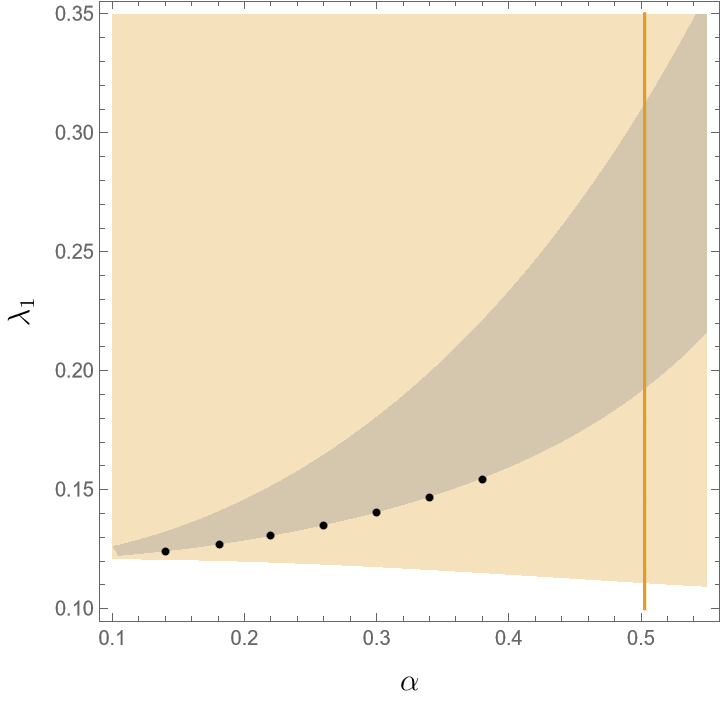}
\caption{Two-dimensional section of the parameter space. {\it Left panel:} All the couplings other than $\alpha$ and $\lambda_1$ are fixed to the values and/or expressions described in the text (in particular, $v_s=150$ GeV, $\mu_3=50$ GeV, $m_{_{Z'}}=250$ GeV and $\lambda_\phi=0.4$, as in Table \ref{table:4}$_1$). In the blue region, tree-level stability and perturbativity of the couplings are realized. In the yellow region, the lowest eigenvalue $m^2_{-}$ in \eqref{mplusminus} is positive. Numerical complications arise when it becomes thinner. The yellow line corresponds to $\sin^2(\alpha-\beta)=0.1$. The red dot indicates the point of coordinates $(\alpha,\lambda_1)=(\beta,\lambda_{\rm SM})$. {\it Right panel}: Same plot as in the left panel, the only difference being the choice for $\lambda_{H\phi}$ and $\lambda_{H'\phi}$, that here are taken as the solution with $\lambda_{H\phi}=\lambda_{H'\phi}$. The black dots indicate the benchmark points in Table \ref{table:last}.} 
\label{fig:parameter space}
\end{figure}

Fixing all the other couplings to the convenient values (and/or expressions) found in the previous sections, the only two free parameters are $\lambda_1$ and $\alpha$. The two-dimensional space they span is restricted by theoretical and phenomenological requirements. Specifically, we impose the tree-level stability conditions \eqref{conditions stability} and \eqref{condition stability 2}, perturbativity of the couplings, positiveness of the eigenvalues \eqref{mplusminus} and an upper bound on the misalignment, namely $\sin^2(\alpha-\beta)\le0.1$, for the latter to be consistent with observations (the precise value of the upper bound is unimportant and was chosen arbitrarily; it only amounts to a shift of the yellow line in the plots shown in Fig.\,\ref{fig:parameter space}). The resulting parameter space is shown in Fig.\,\ref{fig:parameter space} (blue region). The only couplings, other than $\lambda_2$ and $\lambda_{34}$, that depend on $\lambda_1$ and $\alpha$ when adopting the procedure outlined above are  $\lambda_{H'\phi}$ and $\lambda_{H\phi}$, either through the choice $\lambda_{H'\phi}=-2\sqrt{\lambda_2\lambda_\phi}$ in combination with  eq.\,(\ref{semidec couplings}), or through the choice 
\begin{equation}
	\lambda_{H\phi}=\lambda_{H'\phi}=\frac{\mu_3}{\sqrt 2v_\phi}\frac{\sin (\alpha+\beta)}{\cos(\alpha-\beta)}.
\end{equation}
In both cases, the stability and perturbativity constraints on these couplings are milder than those on $\lambda_2$ and $\lambda_{34}$. Nevertheless, the difference between the left panel and the right panel in Fig.\,\ref{fig:parameter space} shows an extreme sensitivity of the sign of $m^2_-$ in \eqref{mplusminus} on the specific choice of the couple $(\lambda_{H\phi}, \lambda_{H'\phi})$ that solves \eqref{semidec couplings}. In the left plot, in fact, the choice $\lambda_{H'\phi}=-2\sqrt{\lambda_2\lambda_\phi}$ leads to a strong constraint on the allowed region of parameter space. In this case, the point $\left(\lambda=\lambda_{\rm SM}, \alpha=\beta \right)$ that represents the tuned and aligned scenario is in a tiny allowed region around which numerical complications arise that make the plot discontinuous. To be certain that this feature is only a result of numerical difficulties, we checked, for several points nearby that are not coloured in yellow, that they correctly have $m^2_->0$. In the right plot, the choice $\lambda_{H\phi}=\lambda_{H'\phi}$ is shown to leave much more freedom to move in the parameter space, as the blue region is entirely contained in the yellow one. Physically, this means that no constraint arises from the condition $m^2_-\ge 0$ in eq.~(\ref{mplusminus}). 

\begin{figure}
	\centering
	\includegraphics[scale=0.5]{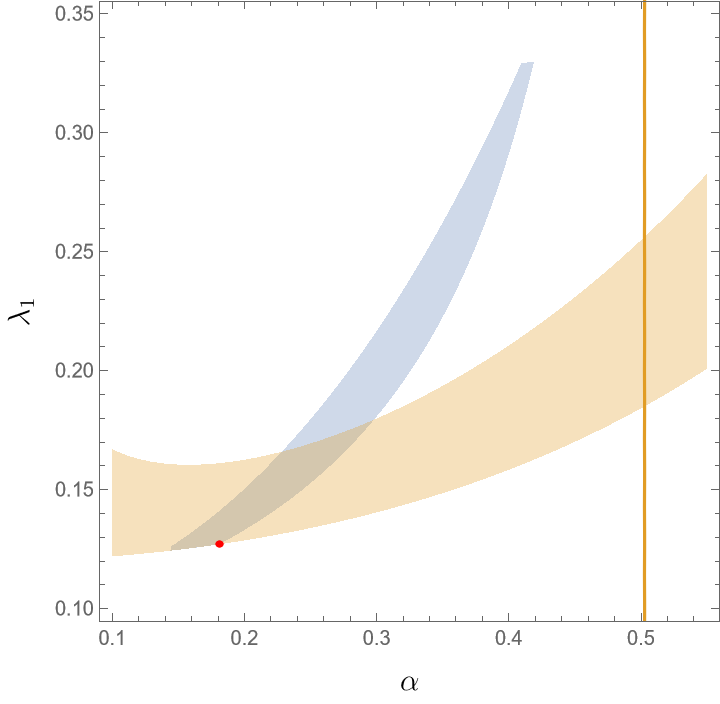}\,\,\,\,\,\,\,\,
	\includegraphics[scale=0.5]{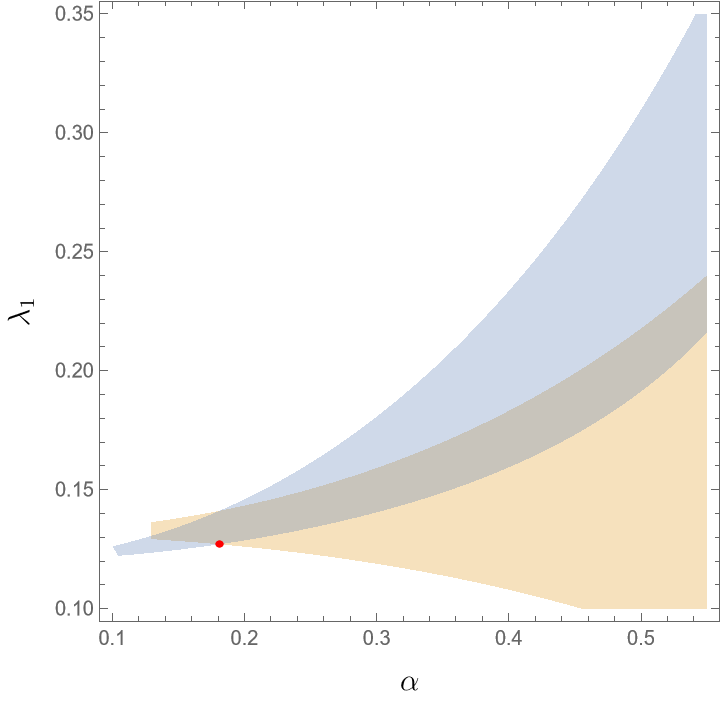}
	\caption{{\it Left panel:} Same plot as in the left panel of Fig.\,\ref{fig:parameter space} for a different choice of $\mu_3$ and $v_\phi$, namely $\mu_3=380$ GeV and $v_s=250$ GeV. The red dot again indicates the point $\alpha=\beta$, $\lambda=\lambda_{\rm SM}$. {\it Right panel:} Comparison between the allowed region of parameter space in our model with the choice $\lambda_{H\phi}=\lambda_{H'\phi}$ (blue region) and the allowed region of parameter space in the corresponding 2HDM (yellow region).}
	\label{fig:parameter space comparison}
\end{figure}

Given the above considerations, we can conclude that, unless specific choices for $\lambda_{H\phi}(\alpha,\lambda_1)$ and $\lambda_{H'\phi}(\alpha,\lambda_1)$ that make their dependence on $\alpha$ and $\lambda_1$ strongly constraining are made, the shape of the allowed region in parameter space does not depend on them and coincides with that of the blue region. As suggested by \eqref{misaligned l2} and \eqref{misaligned l34}, the latter depends on $\mu_3,\, v_\phi$ and $\beta$. To showcase the dependence on the first two parameters, in the left panel of Fig.\,\ref{fig:parameter space comparison} we report the parameter space for a different choice of $\mu_3$ and $v_\phi$. In passing, we note that the above remarks make our results slightly more general and apply, to some extent, to the case of 2HDMs. A comparison between the parameter space in Fig.\,\ref{fig:parameter space} and the one obtained when the scalar singlet is removed (that is, keeping only the constraints on $\lambda_2, \lambda_{34}$ and putting all the parameters related to the scalar singlet to zero) is shown in Fig.\,\ref{fig:parameter space comparison}

\begin{figure}
	\centering
	\includegraphics[scale=0.5]{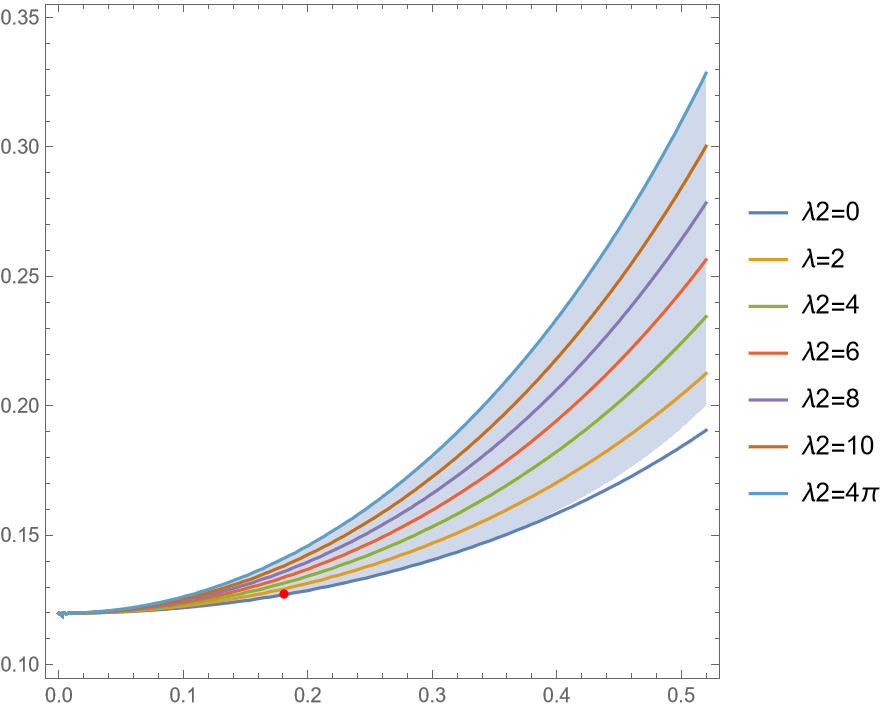} \,\,\,
	\includegraphics[scale=0.5]{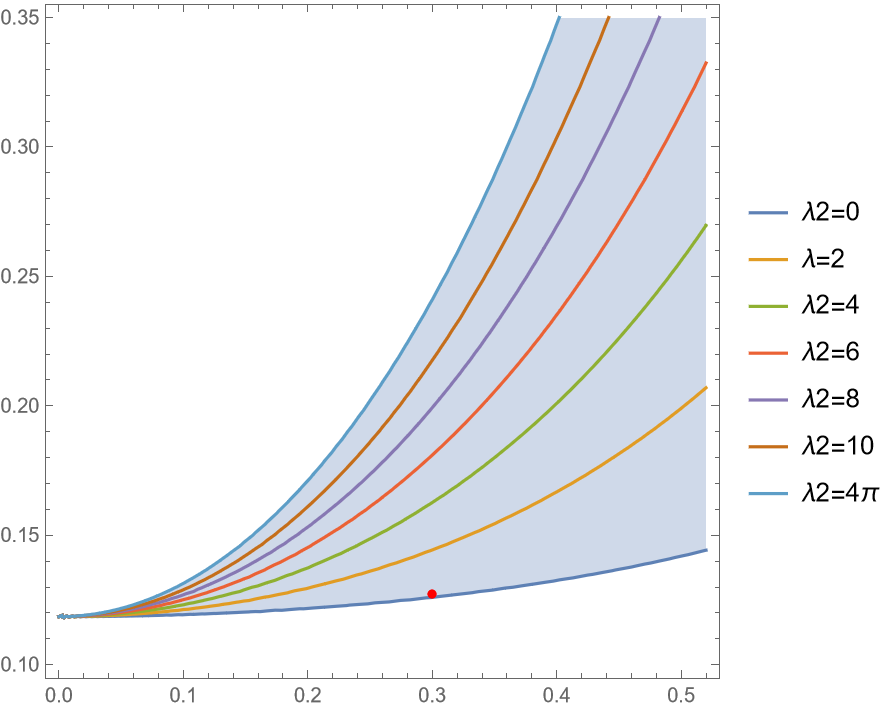}
	\caption{{\it Left panel:} Two dimensional section of the parameter space. The parameters have been chosen in the same way as in the middle panel of Fig.\,\ref{fig:parameter space}. The curves are the curves of constant $\lambda_2$. {\it Right panel}: Same plot as in the left panel for a different choice of $\beta$, namely $\beta=0.3$}
	\label{fig: parameter space l2}
\end{figure}

We performed a numerical investigation of the parameter space with parameters fixed as in the right panel of Fig.\,\ref{fig:parameter space}. As a general property, it is easily understood that at fixed $\lambda_1$ the Landau pole increases and the instability scale decreases as $\alpha$ increases, while at fixed $\alpha$ the Landau pole decreases and the instability scale increases (or disappear for sufficiently low Landau poles) as $\lambda_1$ increases. We found that the competition between these two behaviours is such that on the lower boundary of the allowed region the Landau pole has only very mild variations. This curve of ``constant Landau pole", defined by the equation $\lambda_3+2\sqrt{\lambda_1\lambda_2}=0$ (we note here that it also approximately corresponds to $\lambda_2=0$ for $\alpha\lesssim 0.3$), maximizes the Landau pole, and represents the locus of parameter space where the largest hierarchy between the (in)stability scale and the scale at which perturbation theory becomes unreliable is realized.  Combining these two observations, we can conclude that, as the tuned and aligned scenario (red point in Fig.\,\ref{fig:parameter space}) is itself in the vicinity of the lower boundary, the Landau pole cannot be lifted significantly with respect to that case.

In passing, we also show in Fig.\,\ref{fig: parameter space l2} the curves of constant $\lambda_2$. These curves give us a more comprehensive understanding of the strong sensitivity to the detuning of $\lambda_1$ as a result of their focusing in the small $\alpha$ regime. The smaller the value of $\alpha$, the smaller the distance between the different curves. For $\alpha=\beta$, the relative distance between the point where $\lambda_2=0$ and the point where $\lambda_2=4\pi$ is $\left(\lambda_1^{\lambda_2=4\pi}-\lambda_1^{\lambda_2=0}\right)/\lambda_1^{\lambda_2=0}=4\pi \sin^4\beta/(\lambda_{\rm SM}\cos 2\beta)\simeq0.11$. This relative difference only depends on the angle $\beta$, and thus applies also to the corresponding 2HDM. This can be readily seen in the right panel of Fig.\,\ref{fig:parameter space comparison}, where the vertical extension of the allowed region for $\alpha=\beta$ is the same as in our model. As the distance goes like $\sin^4\beta/\cos2\beta$, it grows for larger values of $\beta$, where the sensitivity to detuning becomes smaller and smaller. This can be seen, for instance, in the right panel of Fig.\,\ref{fig: parameter space l2}. The relative distance in the general, $\alpha\ne \beta$, case is 
\begin{equation}
\delta\lambda\equiv\frac{\lambda_1^{\lambda_2=4\pi}-\lambda_1^{\lambda_2=0}}{\lambda_1^{\lambda_2=0}}=\frac{16 \pi  v^2 \sin ^2\alpha \sin ^2\beta}{\cos2 \alpha \left(4 \lambda_{\rm SM} v^2-\frac{\mu_3 v_\phi}{\sin\beta\cos\beta}\right)+2 \mu_3 v_\phi \cot 2 \beta}.
\end{equation}
The peculiar shape of the allowed region of parameter space shown in Fig.\,\ref{fig:parameter space} is easily understood with the above equation.

We report in Table \ref{table:last} the results obtained for some points on the $\lambda_3+2\sqrt{\lambda_1\lambda_2}=0$ curve. The singularity in the running is always found around $\mu_L\sim 1.57\times 10^9$ GeV. A peculiar feature that can be observed concerns the (in)stability scale. Moving from left to right in the $(\alpha,\lambda_1)$ parameter space, the flow of $\lambda_1(\mu)$ is always positive (that is, for all values of $\mu$ up to the perturbative scale) for $\alpha<\beta$. At some point, while $\alpha$ is still $\alpha<\beta$, the running coupling $\lambda_1(\mu)$ starts to develop an instability scale, that is, a scale where it crosses zero. The point ($\beta, \lambda_{\rm SM}$) is very close from the ``constant Landau pole" boundary, and is in fact found to have this kind of behaviour. This feature eventually disappears for larger values of $\alpha$, where $\lambda_1(\mu)$ is again positive throughout its whole running. The results collected in Table \ref{table:last} also seem to indicate that, when it exists, the instability scale is a convex function of $\alpha$.

\begin{table}[htbp!] 
	\begin{tabular}{|p{1cm}|p{1.1cm}|p{1.1cm}|p{1.3cm}|p{1.1cm}|p{1.0cm}|p{1.0cm}|p{1.6cm}|p{1.6cm}|p{1.6 cm}|}
		\hline
		\multicolumn{10}{|c|}{Benchmark points in Fig.\,\ref{fig:parameter space} (\textit{right})} \\
		\hline
		\centering $\alpha$ & 
		\centering $\lambda_1$ &  \centering $\lambda_2$ & \centering $\lambda_{34}$ & \centering $\lambda_{H\phi}$ & \centering $m_{h_2}$ {\footnotesize(GeV)} & \centering $m_s$ {\footnotesize(GeV)} &  LP {\footnotesize($\times10^9\,$GeV)} &  VSB {\footnotesize($\times10^7\,$GeV)} & Pert {\footnotesize($\times10^8\,$GeV)}
 \\	
		\hline
		0.14 & 0.1241 & $\sim 0$ & 0.3621 & 0.0526 & 219.83 & 178.08 & $1.57$ & --- & $7.22$\\
		$\beta$ & 0.1271 &  $\sim 0$  & 0.2629 & 0.0590 & 220.28 & 178.50 & $1.57$ & $9.60$ & $7.30$ \\
		$0.22$ & 0.1306 & $\sim 0$ & 0.1671 & 0.0651 & 220.87 & 178.95 & $1.58$ & $6.52$ & $7.35$ \\
		$0.26$ & 0.1351 & $\sim 0$  & 0.0672 & 0.0714 & 221.65 & 179.46 & $1.57$ & $12.3$ & $7.36$ \\
		$0.3$ & 0.1405 & 0.0022 & $-0.0352$ & 0.0777 & 222.64 & 180.03 & $1.57$ & $73.7$ & $7.31$ \\
		$0.34$ & 0.1468 & 0.0035  & $ -0.1401$ & 0.0840 & 223.84 & 180.63 & $1.55$ & --- & $7.25$ \\
		$0.38$ & 0.1544 & 0.0375  & $-0.2528$ & 0.0905 & 225.52 & 181.41 & $1.48$ & --- & $6.76$ \\
		\hline
	\end{tabular}
	\caption{Landau pole, stability scale and perturbativity scale for the  benchmark points in the right plot of  Fig.\,\ref{fig:parameter space}. We chose $\sin\beta=0.18$. }
	\label{table:last}
\end{table}

This leads us to the inevitable conclusion that if our model is to explain the muon $g-2$ experimental anomaly, and possibly also the $W$ mass one, it must necessarily be accompanied by new physics that must show up no further than $\mu\sim 10^9$ GeV. A more refined analysis might lead to a more precise definition of the upper bound for the appearance of new physics, but is not expected to change it much.

\section{Conclusions}

We presented in this work new results on the RG flow towards high energies in lepton portal models where both the muon $g-2$ and the $W$ boson experimental anomalies can be explained simultaneously. We found that the choice of parameters favored by the SM anomalies results in unacceptably low Landau pole scales at $\mu_L\sim 10-100$ TeV in the alignment limit for the quartic couplings in the extended Higgs sector. Making a thorough analysis in the Higgs sector with more general relations, we showed that the Landau pole scale $\mu_L$ can be lifted up to $\mu_L\gtrsim 10^9$ GeV, and identified a preferred region of parameter space that maximizes $\mu_L$ while still fitting the experimental anomalies. Such a region of parameter space can potentially ameliorate the stability of the scalar potential. We also showed that, in the cases considered, stringent constraints arise from the Higgs data for scalar couplings and parameters in general scenarios that realize a detuning of the Higgs quartic couplings and/or a misalignment of the two Higgs doublets.    	

As a by-product if our analysis, we learned that if our model is to explain the experimental anomalies, it must be accompanied by new physics that must necessarily appear at scales $\mu\lesssim 10^9$ GeV. Concerning stability, this means that, if it couples strongly enough to the Higgs doublet $H$, such new physics could potentially deviate the RG flow of $\lambda_1(\mu)$ before the latter becomes negative. A final verdict on the existence of a vacuum deeper than the EW one in our model can then only be obtained through the knowledge of its $\mu\sim 10^9$ GeV completion. Our results on the flow of $\lambda_1(\mu)$ should thus be taken as indicating regions of parameter space where the quartic coupling can be lifted to positive values with no need for additional UV physics to do it. A complete study of stability within our model should also contemplate other possibilities, such as the possibility that the running coupling is lifted by additional states, or that the EW vacuum is a metastable one with a sufficiently long lifetime. In both cases, knowledge of the UV physics completing the model is necessary to draw any conclusion.

\section*{Acknowledgments}

The work is supported in part by Basic Science Research Program through the National Research Foundation of Korea (NRF) funded by the Ministry of Education, Science and Technology (NRF-2022R1A2C2003567) and JSPS KAKENHI Grant Number JP24K17040 (KY).

\vspace{1.5cm}

\noindent
{\Large \bf Appendices}

\appendix

\renewcommand{\theequation}{A.\arabic{equation}}

\setcounter{equation}{0}

\section{Mass matrices}
\label{Appendix: A}

We present the effective mass matrices for scalars, gauge bosons and fermions in our model.

\subsection{Scalar masses}

The $\left(h_1,h_2,s\right)$ background of \eqref{parametrization H1}-\eqref{parametrization phi} is unfit to the extraction of Renormalization Group Equations, as both the operators $|H|^2|H'|^2$ and $|H^\dagger H'|^2$ reduce to $h_1^2 h_2^2$ in it, and, when applying the background field method, there would be no way to differentiate the two. For the determination of mass matrices and RGEs, it is convenient to consider, for the two Higgs doublets, the slightly more general background configuration  
\begin{equation}
	H=\frac{1}{\sqrt 2}\left(\begin{array}{c} 0  \\ h_1 \end{array} \right), \qquad H'=\frac{1}{\sqrt 2}\left(\begin{array}{c} h_3  \\ h_2 \end{array} \right).
\end{equation} 
On this background, in fact, $|H|^2|H'|^2=h_1^2(h_2^2+h_3^2)/4$, while $|H^\dagger H'|=h_1^2h_2^2/4$. Contributions that renormalize the coupling $\lambda_3$ can then be distinguished from contributions that renormalize the coupling $\lambda_4$. At the end of the calculation, that is after Renormalization Group Equations have been extracted, one is free to consider the simpler background configuration with $h_3=0$. 

Expanding the potential \eqref{scalarpot} around the $(h_1,h_2,h_3, s)$ background with parametrization 	\begin{eqnarray}
	H&=&\frac{1}{\sqrt 2}\left(\begin{array}{cc} \phi_1+ i \phi_2 \\h_1+\rho_1+i\eta_1 \end{array} \right), \\
	H'&=&\frac{1}{\sqrt 2}\left(\begin{array}{cc} h_3+\phi_3+i\phi_4 \\ h_2+\rho_2+i\eta_2 \end{array} \right),\\
	\phi&=&\frac{1}{\sqrt{2}}(s+\rho_3+i\eta_3).
\end{eqnarray}
we find a $10\times10$ mass matrix. Arranging its entries in the order $(\rho_1,\rho_2,\rho_3,\eta_1,\eta_2,\eta_3,\phi_1,\phi_2,\phi_3,\phi_4)$, it reads  
\begin{eqnarray}
	\label{scalar mass matrix}
	{\cal M}^2_S =\left(\begin{array}{cccccccccc} a & b & d & 0 & 0 & 0 & \alpha & 0 & \beta & 0 \vspace{0.2cm} \\ b & c & e & 0 & 0 & 0 & \gamma & 0 & \delta & 0 \vspace{0.2cm}   \\ d & e  &  f & 0 & 0 & 0 & \eta & 0 & \theta & 0 \\ 
		0 & 0 & 0 & a' & b' & d' & 0 & \zeta & 0 & 0 \vspace{0.2cm}   \\
		0 & 0 & 0 & b' & c' & e' & 0 & \xi & 0 & 0 \vspace{0.2cm}   \\
		0 & 0 & 0 & d' & e' & f' & 0 & \psi & 0 & 0 \vspace{0.2cm}   \\
		\alpha & \gamma & \eta & 0 & 0 & 0 & A & 0 & C & 0 \vspace{0.2cm}   \\
		0 & 0 & 0 & \zeta & \xi & \psi & 0 & B & 0 & E \vspace{0.2cm}   \\
		\beta & \delta & \theta & 0 & 0 & 0 & C & 0 & D & 0 \vspace{0.2cm}   \\
		0 & 0 & 0 & 0 & 0 & 0 & 0 & E & 0 & F \vspace{0.2cm}   \\ \end{array} \right),
\end{eqnarray}
with
\bea
	a&=& \mu^2_1+3\lambda_1 h^2_1 +\frac{\lambda_3}{2} \left(h^2_2 +h_3^2\right)+\frac{\lambda_4}{2} h_2^2+\frac{\lambda_{H\phi}}{2} s^2, \\
	b &=&  \left(\lambda_3+\lambda_4\right)h_1 h_2-\frac{\mu_3}{\sqrt{2}} s, \\ 
	c &=& \mu^2_2+3\lambda_2 h^2_2+\lambda_2 h_3^2+ \frac{\lambda_3 +\lambda_4}{2}h^2_1 +\frac{\lambda_{H'\phi}}{2} s^2, \\
	\label{M2S 13}
	d &=& \lambda_{H\phi}h_1 s -\frac{\mu_3}{\sqrt{2}} h_2, \\ 
	\label{M2S 23}
	e &=& \lambda_{H'\Phi}h_2 s-\frac{\mu_3}{\sqrt{2}}  h_1, \\
	f &=& \mu^2_\phi +3\lambda_\phi s^2+\frac{\lambda_{H\phi}}{2} h^2_1 + \frac{\lambda_{H'\phi}}{2} \left(h^2_2+h_3^2\right), 
\eea	
\bea
	a' &=&\mu^2_1 +\lambda_1 h^2_1  + \frac{\lambda_3}{2} \left(h^2_2+h_3^2\right)+\frac{\lambda_4}{2}h_2^2+\frac{\lambda_{H\phi}}{2}s^2, \\
	b' &=& - \frac{\mu_3}{\sqrt{2}} s, \\ 
	c' &=& \mu^2_2 +\lambda_2 \left(h^2_2 +h_3^2\right)+ \frac{\lambda_3 +\lambda_4}{2}h^2_1 +\frac{\lambda_{H'\phi}}{2} s^2, \\
	d' &=& -\frac{\mu_3}{\sqrt{2}} h_2, \\
	e'&=& \frac{\mu_3}{\sqrt{2}} h_1, \\
	f'&=& \mu^2_\phi + \lambda_\phi s^2 +\frac{\lambda_{H\phi}}{2} h^2_1 + \frac{\lambda_{H'\phi}}{2} \left(h^2_2+h_3^2\right), 
\eea
\bea	
	A&=& \mu^2_1+\lambda_1 h^2_1 +\frac{\lambda_3}{2} \left(h^2_2+h_3^2\right)+\frac{\lambda_4}{2}h_3^2 + \frac{\lambda_{H\phi}}{2} s^2\,=B, \\
	C&=& \frac{\lambda_4}{2}h_1 h_2 - \frac{\mu_3}{\sqrt{2}} s\,=E, \\
	D&=& \mu^2_2+\lambda_2 h^2_2+3\lambda_2 h_3^2 +\frac{\lambda_3}{2} h^2_1+ \frac{\lambda_{H'\phi}}{2} s^2, \\
	F&=& \mu^2_2+\lambda_2 \left(h^2_2+ h_3^2\right) +\frac{\lambda_3}{2} h^2_1+ \frac{\lambda_{H'\phi}}{2} s^2, 
\eea
\bea	
	\alpha&=& \frac{\lambda_4}{2} h_2 h_3, \qquad \beta=\lambda_3 h_1 h_3, \qquad \gamma=\frac{\lambda_4}{2} h_1 h_3,\\
	\delta&=& 2\lambda_2 h_2 h_3, \qquad \eta=-\frac{\mu_3}{\sqrt 2}h_3, \qquad \theta=\lambda_{H'\phi} h_3 s, \\
	\zeta&=&\frac{\lambda_4}{2} h_2 h_3, \qquad \xi=-\frac{\lambda_4}{2} h_1 h_3, \qquad \psi=-\frac{\mu_3}{\sqrt 2} h_3.  
\eea

When $h_3=0$, the Higgs-like, Goldstone-like and charged scalars decouple, and three different mass matrices can be individuated for each of them, as usual. In the text, we denote with $\mathcal M^2_H$ the upper $3\times 3$ matrix of the Higgs-like excitations.

\subsection{Gauge masses}

Expanding the covariant derivatives around the background, we we find the following $5\times 5$ mass matrix in the $(W^1_\mu, W^2_\mu, B_\mu, W^3_\mu, Z'_\mu)$ (in the unitary gauge)
\begin{eqnarray}
	\label{gauge mass matrix}
	{\cal M}^2_V = \left(\begin{array}{ccccc} \frac{g^2}{4}h_{123}^2 & 0 & \frac{g g_Y}{2} h_2 h_3 & 0 & 2 g g_{_{Z'}} h_2 h_3\vspace{0.2cm} \\
		0 & \frac{g^2}{4}h_{123}^2 & 0 & 0 & 0 \vspace{0.2cm} \\
		\frac{g g_Y}{2} h_2 h_3 & 0 &\frac{g^2_Y}{4} h_{123}^2 & \frac{g g_Y}{4} \left(h_3^2-h^2\right) & g_Y g_{_{Z'}} \left(h^2_2+h_3^2\right) \vspace{0.2cm} \\ 0 & 0 & \frac{g g_Y}{4} \left(h_3^2-h^2\right) & \frac{g^2}{4} h_{123}^2 &  g g_{_{Z'}} \left(h_3^2- h^2_2\right)  \vspace{0.2cm} \\ 2 g g_{_{Z'}} h_2 h_3 & 0 &  g_Y g_{_{Z'}} \left(h^2_2+h_3^2\right) &   g g_{_{Z'}} \left(h_3^2-h^2_2\right) & 4g^2_{Z'}(s^2+h^2_2+h_3^2) \end{array}\right),
\end{eqnarray}
with $h^2\equiv h^2_1+h^2_2$, $h_{123}^2\equiv h^2_1+h^2_2+h_3^2$. It is immediate to verify that, when $h_3=0$, $W^1$ and $W^2$ decouple, giving rise to the usual $W$ mass, while the other three components form a $3\times 3$ mass matrix that we refer to with the symbol $\mathcal M_Z^2$ in the text. 

\subsection{Fermion masses}

When $h_3\ne 0$, a fermion mass matrix for the muon, muon neutrino and the VLL can be written. The right-handed neutrino serves only to write a $3\times3$ matrix and can be safely ignored later, especially when we deal with the background considered in the text, where $h_3=0$.   

We consider the mass terms for the lepton doublet and the vector-like lepton as
\begin{eqnarray}
	{\cal L}_{L,{\rm mass}}&=& -M_E {\bar E}E-m_0 {\bar e}e-( m_R {\bar E}_L e_R+m_L {\bar e}_L E_R+m_\nu \bar\nu_l E_R+ {\rm h.c.}) \nonumber \\
	&=&- \left(\bar\nu_L, {\bar e}_L, {\bar E}_L\right) {\cal M}_L  \left(\begin{array}{c} \nu_R \\ e_R  \\ E_R \end{array} \right) +{\rm h.c.}
	\label{leptonmass0}
\end{eqnarray}
where 
\begin{eqnarray}
	{\cal M}_L= \left(\begin{array}{ccc} 0 & 0 & m_\nu\\
		0 &  m_0 & m_L \\ 0 & m_R & M_E \end{array} \right).
\end{eqnarray}
Here, $m_0$ is the bare lepton mass $m_0=\frac{1}{\sqrt{2}} y_l h_1$, $m_R$ and $m_L$ are the mixing masses given by $m_R=\frac{1}{\sqrt{2}}\lambda_E s$ and $m_L=\frac{1}{\sqrt{2}} y_E h_2$, respectively, and $m_\nu=\frac{1}{\sqrt{2}} y_E h_3$.
The squared mass matrix is thus 

\begin{eqnarray}
	\label{fermion mass matrix}
	\mathcal M^2_f\equiv \mathcal M^\dagger_L\mathcal M_L=\left(\begin{array}{ccc} 0 & 0 & 0 \\
		0 & m_0^2+m_R^2 & m_0m_L+m_RM_E \\ 0 & m_0m_L+m_RM_E & m_\nu^2+ m_L^2+M_E^2 \end{array} \right).
\end{eqnarray}

Finally, the top quark mass is $m_t=\frac{y_t}{\sqrt 2} h_1$.

	\setcounter{equation}{0}
	\renewcommand{\theequation}{B.\arabic{equation}}
\section{Renormalization Group Equations}
\label{Appendix: B}

We present the RG equations for the parameters in the scalar potential, gauge and Yukawa couplings in the model.

\subsection{Extraction of the scalar RGEs from the potential}
Inserting the explicit expressions \eqref{V0} and \eqref{V1} in \eqref{RGE potential} we obtain the following system of equations ($k$ is the running scale, the symbols $\beta_{\lambda_i}$ and $\gamma_{\mu_i}$ indicate $k\frac{d}{d k}\lambda_i$ and $k\frac{d}{d k}\mu_i$, respectively, $\gamma_i$ are the anomalous dimensions and $\alpha_i$ are the coefficients of the quadratically UV-sensitive terms defined in \eqref{quad UV-sensitive}):
	\begin{align}
	\mu_1^2\gamma_{\mu_1^2}-2\gamma_1 \mu_1^2=& -2\alpha_1 k^2+\frac{6 \lambda_1\mu_1^2+2\left(\lambda_3+\lambda_4\right) \mu_2^2+\lambda_{H\phi}\mu_\phi^2+\mu_3^2}{8 \pi ^2}, \\
	\mu_2^2\gamma_{\mu_2^2}-2\gamma_2 \mu_2^2=& - 2\alpha_2 k^2+\frac{6 \lambda_2\mu_2^2+2\lambda_3 \mu_1^2+\lambda_4\mu_1^2+\lambda_{H'\phi}\mu_\phi^2+\mu_3^2-2y_E^2M_E^2}{8 \pi ^2},\\
	\mu_\phi^2\gamma_{\mu_\phi^2}-2\gamma_\phi \mu_\phi^2=& - 2\alpha_3 k^2+\frac{4 \lambda_\phi\mu_\phi^2+2\lambda_{H\phi} \mu_1^2+2\lambda_{H'\phi}\mu_2^2+2\mu_3^2-2\lambda_E^2M_E^2}{8 \pi ^2}, \\
	\beta_{\lambda_1}-4\gamma_1\lambda_1=&\frac{9 g^4+6 g^2g_Y^2+3g_Y^4}{128 \pi ^2}+\frac{3 \lambda_1^2}{2 \pi ^2}+\frac{2\lambda_3(\lambda_3+\lambda_4)+\lambda_4^2}{16 \pi ^2}+\frac{\lambda_{H\phi}^2}{16 \pi ^2}-\frac{3 y_t^4+y_l^4}{8 \pi ^2},\\
	\beta_{\lambda_2}-4\gamma_2\lambda_2=&\frac{9 g^4+6 g^2g_Y^2+3g_Y^4}{128 \pi ^2}+\frac{3g_{_{Z'}}^2(g^2+g_Y^2)+24g_{_{Z'}}^4}{4\pi^2}+\frac{3 \lambda_2^2}{2 \pi ^2}\nonumber \\&+\frac{2\lambda_3(\lambda_3+\lambda_4)+\lambda_4^2}{16 \pi ^2}
	+\frac{\lambda_{H'\phi}^2}{16 \pi ^2}-\frac{y_E^4}{8 \pi ^2},\\
	\beta_{\lambda_\phi}-4\gamma_\phi\lambda_\phi=&\frac{6g_{_{Z'}}^4}{\pi^2}+\frac{5 \lambda_\phi^2}{4 \pi ^2}+\frac{\lambda_{H\phi}^2}{8 \pi ^2}+\frac{\lambda_{H'\phi}^2}{8 \pi ^2}-\frac{\lambda_E^4}{8 \pi ^2},\\
	\beta_{\lambda_3}-2(\gamma_1+\gamma_2)\lambda_3=&\frac{9 g^4-6 g^2g_Y^2+3g_Y^4}{64 \pi ^2}+\frac{(\lambda_1+\lambda_2)(6\lambda_3+2\lambda_4)+2\lambda_3^2+\lambda_4^2}{8 \pi ^2}+\frac{\lambda_{H\phi}\lambda_{H'\phi}}{8 \pi ^2},\\
	\beta_{\lambda_4}-2(\gamma_1+\gamma_2)\lambda_4=&\frac{3 g^2g_Y^2}{16 \pi ^2}+\frac{\lambda_4(\lambda_1+\lambda_2+2\lambda_3+\lambda_4)}{4 \pi ^2}-\frac{y_E^2y_l^2}{4\pi^2},\\
	\beta_{\lambda_{H\phi}}-2 (\gamma_1 +\gamma_\phi)\lambda_{H\phi}=&\frac{2\lambda_{H\phi}^2+6\lambda_1\lambda_{H\phi}+4\lambda_\phi\lambda_{H\phi}+2\lambda_3\lambda_{H'\phi}+\lambda_4\lambda_{H'\phi}}{8 \pi ^2}-\frac{\lambda_E^2 y_l^2}{4 \pi ^2}, \\
	\beta_{\lambda_{H'\phi}}-2 (\gamma_2 +\gamma_\phi)\lambda_{H'\phi}=&\frac{12 g_{_{Z'}}^4}{\pi ^2}+\frac{2\lambda_{H'\phi}^2+6\lambda_2\lambda_{H'\phi}+4\lambda_\phi\lambda_{H'\phi}+2\lambda_3\lambda_{H\phi}+\lambda_4\lambda_{H\phi}}{8 \pi ^2},\\
	\beta_{\mu_3}-(\gamma_1+\gamma_2+\gamma_\phi) \mu_3=&\frac{\lambda_3+2\lambda_4+\lambda_{H\phi}+\lambda_{H'\phi}}{8 \pi ^2}\mu_3+\frac{y_E \lambda_E y_l M_E}{4\pi^2}.
\end{align}   
The equations above are Wilsonian RG equations in the UV regime of the flow defined as $k^2\gg m_i(k)^2$, with $m_i(k)$ the physical running masses. Subtracting the terms proportional to $k^2$ from the first three equations, that corresponds to performing a fine-tuning in perturbation theory and putting the system close to the critical surface of the Gaussian fixed point in the broader RG context \cite{Branchina:2022gll}, the usual perturbative RGEs are found. 

The anomalous dimensions can be calculated from the two-point functions of $h_1$, $h_2$ and $s$ as 
\begin{equation}
	\gamma_i=-\frac12\Lambda\frac{\partial}{\partial\Lambda}\left(\frac{\partial}{\partial p^2} \Sigma(p^2)\right)_{p^2=m_i^2}.
\end{equation} 
As it is well-known, the only diagrams that contribute at the one-loop level are fermionic sunset diagrams and mixed gauge-Goldstone sunset diagrams (pure gauge diagrams can also give $p^2$ contributions that are however suppressed by powers of $\Lambda$, so they do not contribute to the beta function). 
The results for anomalous dimensions are 
\begin{equation}
	\begin{cases}
		\gamma_1=-\frac{1}{16\pi^2}\left(\frac{3}{4}\left(3g^2+g_Y^2\right)-3y_t^2-y_l^2\right), \\
		\gamma_2=-\frac{1}{16\pi^2}\left(\frac{3}{4}\left(3g^2+g_Y^2\right)+12g_{_{Z'}}^2-y_E^2\right),\\
		\gamma_\phi=-\frac{1}{16\pi^2}\left(12g_{_{Z'}}^2-\lambda_E^2\right).
	\end{cases}
\end{equation}

\subsection{Gauge RGEs}

The general equation for the non-abelian gauge RGEs is
\begin{equation}
	\beta_g=-\frac{g^3}{16\pi^3}\left(\frac{11}{3}C_2(G)-\frac{4}{3}n_f C(r)-\frac 13 n_s C(r)\right)
\end{equation}
where $C_2(G)=N$ and $C(r)=1/2$ for $SU(N)$. 
\begin{itemize}
	\item For $SU(3)$ we have, as in the SM, $n_f=6$, $n_s=0$;
	\item For $SU(2)$ we have the same DOFs as in the SM plus the second Higgs doublet, so that $n_f=6$ and $n_s=2$.
\end{itemize}
For $U(1)$ groups we have in general 
\begin{equation}
	\beta_g=\frac{g^3}{12\pi^2}\left(\sum_i n^i_f (q^i_f)^2+\frac{1}{4}n_s^i(q^i_s)^2\right).
\end{equation}
where $q^i$ are the charges under the gauge group. 

The straightforward application of these equations to our model leads to the following RGEs for the gauge couplings
\begin{equation}
	\begin{cases}
		\beta_{g_s}=-\frac{7}{16\pi^2}g_s^3, \\
		\beta_{g}=-\frac{3}{16\pi^2}g^3, \\
		\beta_{g_{_Y}}=\frac{25}{48\pi^2}g_Y^3, \\
		\beta_{g_{_{Z'}}}=\frac{7}{12\pi^2}g_{_{Z'}}^3
	\end{cases}
\end{equation}
where $g_s$ is the strong coupling, $g$ the $SU(2)$ coupling, $g_{_Y}$ the hypercharge one and $g_{_{Z'}}$ the $U(1)'$ one.

\subsection{Fermion RGEs}

The well-known RGE for the top Yukawa coupling is 
\begin{equation}
	\beta_{y_t}=\frac{y_t}{16\pi^2}\left(\frac{9}{2} y^2_t - 8 g^2_3 - \frac{9}{4}g^2 -\frac{17}{12} g^{\prime 2} \right).
\end{equation}
Calculating the relevant diagrams, we find for the two other Yukawa couplings the following set of RGEs
\begin{equation}
	\begin{cases}
		\beta_{y_E}=\frac{y_E}{16\pi^2}\left(\frac{5}{2}y_E^2+\frac{y_l^2}{2}-\frac{15}{4}g_Y^2-\frac{9}{4}g^2-12g_{_{Z'}}^2\right) \nonumber\\
		\beta_{\lambda_E}=\frac{\lambda_E}{16\pi^2}\left(2\lambda_E^2+y_l^2-6g_Y^2-12g_{_{Z'}}^2\right).
	\end{cases}
\end{equation} 
The last two equations needed to close the full set of RGEs are those for $M_E$ and $y_l$. They are easily found and read
\begin{equation}
	\begin{cases}
		\gamma_M=\frac{3g_Y^2+12g_{_{Z'}}^2}{8\pi^2}\\
		\beta_{y_l}=\frac{y_l}{16\pi^2}\left(\frac{y_E^2+\lambda_E^2}{2}+3y_t^2+\frac32y_l^2-\frac{9}{4}g^2-\frac{15}{4}g_Y^2\right)
	\end{cases}
\end{equation}
where $\gamma_M=\frac{k}{M}\frac{\partial M}{\partial k}$.

\end{document}